\documentclass{aastex701}
\usepackage{amsmath,amsfonts,amssymb}
\usepackage{amsthm}
\usepackage{bbm}
\usepackage{longtable}
\usepackage{booktabs}
\usepackage{graphicx}
\usepackage{mathrsfs}
\usepackage{booktabs}

\usepackage{caption}
\usepackage{enumitem}
\usepackage{subfigure}

\usepackage{float}
\usepackage{emptypage} 
\usepackage{fancyhdr}
\usepackage[marginal]{footmisc}
\usepackage{subfigure}
\usepackage{listings}
\usepackage{xcolor}
\usepackage{array}
\usepackage{algorithm}
\usepackage{algpseudocode}

\begin{document}

\title{Jovian Polar Vortex Generation in a Shallow-Water Model with Quasi-Stochastic Convection}

\author[orcid=0009-0007-7630-953X,sname='Ruan']{Yuhang Ruan}
\affiliation{Nanjing--Helsinki Institute in Atmospheric and Earth System Sciences, Nanjing University, Suzhou, Jiangsu 215163, China}
\email{yuhang\_ruan@smail.nju.edu.cn}

\author[orcid=0000-0002-9524-0001,sname='Fu']{Hao Fu}
\affiliation{School of Atmospheric Sciences, Nanjing University, Nanjing, Jiangsu 210023, China}
\email{haofu@nju.edu.cn}

\author[orcid=0000-0002-8280-3119,sname='Li']{Cheng Li}
\affiliation{Department of Climate and Space Sciences and Engineering, University of Michigan, Ann Arbor, MI 48109, USA}
\email{chengcli@umich.edu}

\author[sname='Yang']{Da Yang}
\affiliation{Department of Geophysics, Stanford University, Stanford, CA 94305, USA}
\email{dayang@stanford.edu}

\correspondingauthor{Hao Fu (\href{mailto:haofu@nju.edu.cn}{haofu@nju.edu.cn})}

\begin{abstract}
The mechanisms governing the formation and maintenance of polar vortices on Saturn and Jupiter represent a significant challenge in current planetary atmospheric dynamics. In the presence of internal heat sources, the existence of condensable components leads to widespread moist convection in the polar regions of these gas giants. However, the role of moist convective organization in the formation and maintenance of such polar vortices remains poorly understood. To investigate the interaction between convection and polar vortices, this study develops a polar $\gamma$-plane shallow-water model coupled with a quasi-stochastic convective parameterization. In this quasi-stochastic scheme, convection is prescribed as localized impulsive mass sinks, with the inception probability depending on local vorticity, thereby assigning a higher convective likelihood to cyclonic regions. Numerical experiments successfully reproduce shielded cyclones and persistent vortex crystals reminiscent of Jupiter's polar configuration. Increasing mesoscale feedback promotes vortex growth, shield formation, and spectral organization, while the characteristic vortex scale depends on both the deformation radius and the convective parameterization. The resulting patterns exhibit both clockwise and counterclockwise structural drift, whose direction and rate are not adequately captured by a conventional Rossby wave phase-speed estimate. A Biot--Savart diagnostic reveals competing contributions to structural drift, with predominantly counterclockwise advection induced by the other vortices and clockwise advection induced by the background vorticity outside the identified vortices. These results provide a possible formation pathway for shielded vortex crystals and highlight the role of convective feedback in vortex formation.
\end{abstract}


\section{Introduction}\label{section:1}
Over the past two decades, the Cassini and Juno missions have unveiled rich polar dynamics on giant planets. Among the many observations of the polar regions, the most striking phenomenon is the polar vortex, especially that of Jupiter. As shown in Fig.~\ref{fig:convection_scheme}, polar vortices exhibit different patterns on different planets. On Saturn, there is a single polar vortex at both the north and south poles, and the north one is surrounded by jet streams. On Jupiter, the situation is different. We can see that in addition to a single vortex near the poles, there is also a ring of ``satellite" vortices (also known as circumpolar cyclones, CPCs) around it. The main energy source of such vortices is considered to be internal heat because Jupiter has almost zero obliquity \citep{Fletcher2008,Li2018}. Such internal heat and condensable substances give rise to moist convection and storms in their atmosphere, which are thought to drive cyclones and jets \citep{Li2006, Ingersoll2020, Siegelman2022-2, Cai2026}.

%There is also strong evidence for polar vortices on ice giants \citep{LUSZCZCOOK2010938,KARKOSCHKA2015294,Akins2023}. 

%Several methods are used to model vortices. 

Many previous studies have focused on reproducing polar vortices on Jupiter or Saturn with numerical models. \citet{Donnelly} use Saturn-DYNAMICO GCM to model the whole Saturn atmosphere, including giant vortices, \citet{Yadav2020} use MagIC to simulate both Jupiter and Saturn's rotating convection on shell, \citet{Siegelman2022-1} and \citet{Shi2026} solved Quasi-Geostrophic Potential Vorticity (QGPV) equation on polar cap, while \citet{ONeill2016}, \citet{Li2020}, \citet{Brueshaber2019}, \citet{Brueshaber2021} and \citet{CHEN2025116438} solved shallow water (SW) equations on polar cap. There is also diversity among the works on the shallow water model. \citet{ONeill2016} used a 2.5-layer SW model and included mature cyclones within the domain, whereas storms are treated as a random force in multistorm simulations. They emphasize the role of the nondimensional beta parameter ($\hat\beta=a/L_D$), which is equivalent to the planetary Burger number ($Bu_p=(L_D/a)^2$), and total energy density in the formation of a strong polar vortex, where $L_D$ is deformation radius and $a$ is the planetary radius. \citet{Brueshaber2019} and \citet{Brueshaber2021} used 1.5-layer SW models, where they put storms into the domain one by one, which is seen as a turbulent force, and divided the dynamical conditions into several schemes. Although \citet{Siegelman2022-1} and \citet{CHEN2025116438} use different equations, they shared the similar initial condition, i.e., the vortex perturbation is randomly seeded and uniformly distributed in direction.

\citet{Li2020} also used a 1.5-layer SW model to reveal the role of the shield around the polar cyclones, where the cyclones' configuration are initially given. They found that the negative-vorticity shield around a cyclonic core can stabilize the vortex pattern in some conditions. This is consistent with \citet{Stellmach2014}'s finding, where they found that intense vortex cores shielded by weaker sheets of opposite vorticity in the direct numerical modeling of rapidly rotating Rayleigh-Bénard convection can prevent them from being destroyed by vortex-vortex interactions. Although \citet{Siegelman2022-1} mentioned that the shields are not necessary for maintaining stable vortex patterns, \citet{CHEN2025116438} subsequently identified signatures of shielding in the PV profiles of those simulations and emphasized its role in suppressing vortex mergers. Observations have also revealed shielded vortices at Jupiter's north pole \citep{ingersoll2022vorticity}. It is therefore worthwhile to investigate how shielded vortices emerge under turbulent forcing.

Research on gas giants' polar vortices has extended to comparative studies with vortices on other planets, especially on terrestrial planets. Saturn's south polar vortex shares similarities with tropical cyclones on Earth, including cyclonic rotation, a warm core, and a cloudy eyewall \citep{Dyudina2009}. Thus, terrestrial tropical cyclone (TC) models are also worth considering for gas-giant researchers. SW model has also been extensively used in TC research, as reviewed below. \citet{Ooyama1969} designed a 3-layer SW model to simulate the life cycle of a TC; \citet{Schecter2009} extended his SW model and reproduced hurricane formation in diabatic Ekman turbulence. and \citet{Garner2023} modeled a stable TC worlds in rotating radiative–convective equilibrium(RRCE) with 3-layer SW model. Recently, \citet{FuONeill2025c} designed a stochastic quasi-2D model to study the formation and interaction of convectively coupled vortices. 

%Such modeling is inspiring for studying polar vortex generation in gas giants.

Stable polar vortex patterns can also undergo collective motion. Observations of Jupiter's polar cyclones show a systematic westward drift, which \citet{gavriel2023} explained using $\beta$-drift and a center-of-mass approach. Laboratory experiments further show that mutual advection can reverse the drift of vortex crystals from westward to eastward as their spacing decreases \citep{BENZEGGOUTA2026119877}. These findings motivate examining the collective rotation of vortex patterns generated and maintained by convection, and its relationship to the vorticity distribution within and around the vortices.

This paper aims to investigate the formation and collective motion of shielded polar vortex patterns using a SW model with a single active layer on a polar $\gamma$ plane. Using quasi-stochastic convective parameterization following \citet{FuONeill2025c}, we examine how vorticity-dependent convective feedback promotes vortex growth, shielding, and organization into persistent vortex crystals. We further diagnose structural drift, defined here as the rotation of the recentered vortex pattern, and use a Biot--Savart diagnostic to relate this motion to velocities induced by the simulated vorticity field.

This paper is organized in the following sequence. Section \ref{Section:2} introduces the shallow water model and our settings, including parameterization, non-dimensional numbers, numerical method, and experimental plan. Section \ref{Section:3} shows the simulation results and analyzes the vortices' drift behavior. In Section \ref{Section:4}, we summarize the paper and discuss the further problems.

%, where convection initiates

\section{Model}\label{Section:2}
\subsection{Shallow Water Equation}
We use a 1.5-layer shallow-water model to represent the lower troposphere over a polar cap. The governing equations are
\begin{align}
\frac{\partial \mathbf{V}}{\partial t}
+(\mathbf{V}\cdot\nabla)\mathbf{V}
+f\hat{\mathbf{k}}\times\mathbf{V}
&=
-g'\nabla h
+\nu\nabla^2\mathbf{V}
+\mathbf{D}_{\rm S}
-\boldsymbol{\Gamma}\mathbf{V}
,
\\
\frac{\partial h}{\partial t}
+\nabla\cdot(h\mathbf{V})
&=
Q_{\rm cool}
+Q_{\rm conv}
-\gamma_hh'
\end{align}
where $\mathbf{V}=(u,v)$ is the horizontal velocity, $h$ is the lower troposphere thickness, $H_0$ is its reference value, $h'=h-H_0$ its perturbation, $g'$ indicates the reduced gravity and $\nu$ is the constant viscosity. $f\hat{\mathbf{k}}\times\mathbf{V}=f(-v,u)$ is the Coriolis term and $\mathbf{D}_{\rm S}$ denotes the divergence of the Smagorinsky subgrid stress. $Q_{\rm cool}$ is the radiative cooling term. We use Newtonian cooling to represent the radiative cooling of the troposphere, which indicates $Q_{\text{cool}}=-\frac{h'}{\tau_{\text{cool}}}$, where $\tau_{\rm cool}$ is the cooling time scale. $Q_{\rm conv}$ is the convective forcing term, which will be introduced later.

The polar variation of the Coriolis parameter is represented by
\begin{equation*}
    f(r)=f_0[1-r^2/(2a^2)],
\end{equation*}
where $a$ is the planetary radius, $f_0$ is the polar Coriolis parameter, and $r$ is the distance from the pole. This is the leading-order expansion of the spherical Coriolis parameter about the pole. The corresponding local planetary-vorticity gradient is $\beta=|d f/d r|=f_0r/a^2$.

To simulate the polar cap within a doubly periodic domain, we adopt a sponge layer boundary condition. We apply strong damping beyond a certain distance from the pole to prevent eddies and waves from reentering the domain at the boundaries. The damping rate of the sponge layer $\gamma$ is 0 at the edge of the disk and increases quadratically until the edge of the domain. The damping rate of each term is determined by $\tau_{i}$. We can express the damping coefficient as
\begin{equation*}
    \gamma_{i}(\lvert\mathbf{x}\rvert)=\tau_{i}^{-1}(\frac{\lvert\mathbf{x}\rvert-R_{\text{D}}}{R_{\rm c}-R_{\text{D}}})^2\mathcal H\!\left(\lvert\mathbf{x}\rvert-R_{D}\right),
\end{equation*}
where $R_{\rm D}$ is the radius of the polar region, which is set to be $0.45L$, and $L=L_x=L_y$ is the length of the square domain; $R_{\rm c}$ is the distance from the pole to a corner of the square domain. We place the pole in the middle of the domain and set $\mathbf x_{\text{pole}}=(0,0)$. $\mathcal H(x)$ is Heaviside function, which means $\mathcal H(x)=0$ if $x<0$ and $\mathcal H(x)=1$ if $x>0$. $- \boldsymbol{\Gamma} \mathbf{V}$ and $-\gamma_{\text{h}}{h'}$ are linear damping terms, where $\boldsymbol{\Gamma}=\begin{pmatrix} \gamma_u & 0 \\ 0 & \gamma_v \end{pmatrix}$ is a diagonal damping tensor.

\subsection{The Quasi-Stochastic Convective Parameterization}
Both \citet{Brueshaber2019} and \citet{FuONeill2025c} treat a single convective event or storm as a localized height-perturbation pulse that drives convergent (or divergent) motion and a corresponding vorticity perturbation. Following this framework, we model the convective events as dynamic mass sink terms, or equivalently, geopotential sink terms. The forcing sign depends on the dynamical response retained by the model. \citet{Li2006} represent mass injection into the overlying weather layer, producing an anticyclonic response, whereas \citet{ONeill2016} represent upward mass transfer through a lower-layer sink and an upper-layer source in a two dimension model. Beyond this layer-specific interpretation, idealized studies show that nonlinear stretching and vertical transport can generate a cyclonic barotropic component without surface friction \citep{FuSun2024,FuDavis2025}. Here, negative thickness pulses retain an effective convergent, cyclonic response to convection in the modeled layer. This is a simplified representation, not a complete description of the vertical redistribution of mass and momentum.

Each convective event's forcing has a Gaussian spatial distribution, and the imposed thickness anomaly follows the rising branch of a Gaussian temporal envelope; correspondingly, the forcing rate is given by its time derivative and is retained only before the peak; see equation~\ref{eq4}. On Earth, the spreading length scale of convective events $R_{\rm conv}$ is affected by cold pools, gravity waves, and anvil clouds’ nonlocal radiative effect on Earth \citep{Mapes1993,Yang2013,Brueshaber2019,FuONeill2025b}. On Jupiter, the synoptic layer has no clear lower boundary and thus no cold pools. The nonlocal radiative effect may be weak, as Jupiter is much cooler than Earth and the Stefan-Boltzmann law suggests its influence is small. Therefore, we consider that the nonlocal triggering of convection by gravity waves is the main factor determining $R_{\rm conv}$.

\begin{figure*}[t]
\centering
\includegraphics[width=\textwidth]{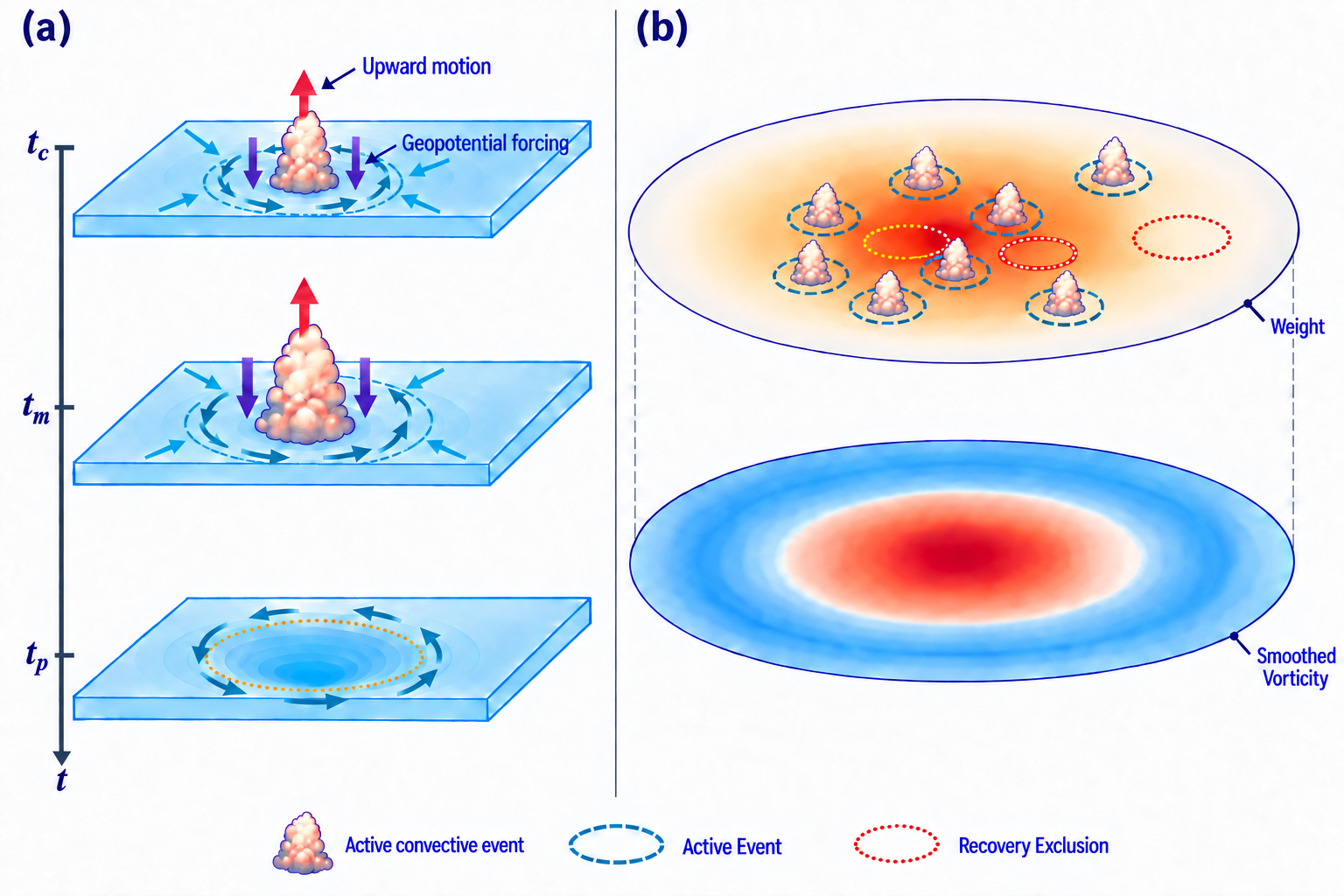}
\caption{Schematic of the quasi-stochastic convective parameterization. (a) An individual event at initiation ($t_c$), maximum forcing ($t_m=t_p-\tau_{\rm conv}/\sqrt{2}$), and termination ($t_p=t_c+2\tau_{\rm conv}$). Purple downward arrows indicate negative geopotential forcing, red upward arrows depict convective ascent, and blue arrows indicate horizontal convergence. The depression illustrates the localized thickness response. (b) Corresponding smoothed-vorticity and initiation-weight fields. Red, white, and blue represent positive, zero, and negative smoothed relative vorticity; deeper orange indicates larger initiation weight. Clouds denote active convective events, dashed circles their exclusion footprints, and dotted circles regions temporarily excluded from further initiation. The two planes represent the same horizontal region. All colors and spatial arrangements are illustrative. Post-event exclusion in (a) represents $c_{\rm cool}>2$.}
\label{fig:convection_scheme}
\end{figure*}

For an individual convective event triggered at time $t_c$ and location $\mathbf{x}_c$, we expect the perturbation to reach its peak negative anomaly magnitude $\Delta h$ at $t_p = t_c + 2\tau_{\text{conv}}$, where $\tau_{\text{conv}}$ is the characteristic duration of the convective event. The total mass forcing field at any given time is the superposition of all active events:
\begin{equation}
Q_{\rm conv}(\mathbf{x},t)=\Delta h
\left[
\sum_{i\in\mathcal S(t)}
\mathcal{T}_i(t)
\delta(\mathbf{x}-\mathbf{x}_{c,i})
\right]
*G_{\rm conv}(\mathbf{x};R_{\rm conv})\,
\mathcal H\!\left(R_D-\lvert\mathbf{x}\rvert\right), \qquad G_{\rm conv}(\mathbf{x};R_{\rm conv})
=
\exp\left(-\frac{|\mathbf{x}|^2}{2R_{\rm conv}^2}\right),
\end{equation}
where $\mathcal{S}$ is the active event set collecting the tuple parameters $(\mathbf{x}_c, t_p, \Delta h, \tau_{\text{conv}})$ for all currently triggered storms,  $\mathcal T_i$ is the time coefficient of the convective event, $\delta$ is the Dirac delta, $*$ denotes the spatial two-dimensional convolution, and $G_{\rm conv}$ is a dimensionless Gaussian footprint with unit peak, so that $\Delta h$ specifies the amplitude of the imposed thickness envelope. Unlike the smoothing kernel $G_{\rm smooth}$, it is not normalized to unit spatial integral; its integral over the plane is $2\pi R_{\rm conv}^2$, which is retained in the mean forcing and linear analysis. The temporal evolution obeys
\begin{equation}
\mathcal{T}_i(t)=
\begin{cases}
-\dfrac{2(t-t_{p,i})}{\tau_{\rm conv}^2}
\exp\!\left[-\dfrac{(t-t_{p,i})^2}{\tau_{\rm conv}^2}\right],
& t_{c,i}\le t<t_{p,i},\\[6pt]
0, & \text{otherwise},
\end{cases}\label{eq4}
\end{equation}
Figure~\ref{fig:convection_scheme}(a) illustrates the localized geopotential forcing and the associated convergent motion during an individual event.

Rather than triggering convection purely randomly, we weight the spatial distribution of convective initiation by the smoothed background relative vorticity. Our convective seeding scheme differs fundamentally from that of \citet{FuONeill2025c}. Their scheme is Lagrangian-based, seeding events at random locations at a given interval. In contrast, our scheme is Eulerian-based, with the convective initiation probability determined locally at each grid point dynamically. As in \citet{Li2006}, initiation responds to the local flow state; here, however, it depends probabilistically on smoothed relative vorticity rather than on a layer-thickness threshold.

The convective distribution function is defined as:

\begin{equation}
w(\mathbf{x},t)
=
\left[
1+\eta\frac{\widetilde{\zeta}(\mathbf{x},t)}{f_0}
\right]_+
\mathcal H\!\left(R_D-\lvert\mathbf{x}\rvert\right),
\qquad \widetilde{\zeta}(\mathbf{x},t) = \zeta(\mathbf{x},t) * G_{\rm smooth}(\mathbf{x};l_g),
\end{equation}
where $[s]_+=\max(s,0)$, $\widetilde{\zeta}$ is the background relative vorticity smoothed by a spatial Gaussian filter of scale $l_g$, and $G_{\rm smooth}(\mathbf{x};l_g)
=
(\pi l_g^2)^{-1}
\exp\left(-|\mathbf{x}|^2/l_g^2\right)
$. The factor $\mathcal H\left(R_D-\lvert\mathbf{x}\rvert\right)$ restricts convection to the physical polar domain, while $\eta$ controls the tendency of convective events to aggregate in regions of positive smoothed relative vorticity.

On Earth, the observed association between cyclonic vorticity and deep convection or lightening can partly arise from surface-friction-driven Ekman pumping \citep{Molinari2004,FuONeill2025c}. This mechanism, however, does not directly apply to giant planets, which lack a solid lower boundary. There, the prescribed vorticity dependence is motivated by both observations and the proposed vertical structure of cyclones. Jovian lightning and moist convective outbreaks preferentially occur in cyclonic-shear regions and cyclonic folded-filamentary regions \citep{LITTLE1999306,Fletcher2026,Brown2018}. On Saturn, lightning has likewise been detected in regions of cyclonic vorticity and within individual cyclones \citep{DYUDINA20131020,GUNNARSON2023115228}. One proposed explanation is that cyclonic pressure anomalies raise deep isentropes and water-rich layers, thereby facilitating the triggering of moist convection \citep{Fletcher2026}. Because this vertical triggering mechanism is not resolved by our 1.5-layer model, we instead use the smoothed relative vorticity as an empirical proxy that biases convective initiation toward cyclonic regions. Within the model, this choice also closes an idealized positive feedback: each convective mass sink drives horizontal convergence and amplifies cyclonic vorticity through stretching, so that $\widetilde{\zeta}$ retains a coarse-grained memory of recent nearby convection. Conversely, the initiation weight decreases in anticyclonic regions and is set to zero when $\widetilde{\zeta}/f_0\leq-1/\eta$. This cutoff ensures a nonnegative probability and represents an idealized suppression of convection rather than a universal physical threshold.

To regulate the timing and frequency of convective events, we introduce the expected global convective frequency $N$ (with a unit of $s^{-1}$). At each time step $\Delta t$, the baseline probability density of a new convective event is formulated as:

\begin{equation*}
    \lambda_{\text{raw}}(\mathbf{x},t) =N  \frac{w(\mathbf{x},t)}{\iint w(\mathbf{x} ,t) d\mathbf{x}}
\end{equation*}

To prevent the unphysical clustering of identical storms in an excessively narrow region—since overlapping convective cells are typically considered a single mesoscale convective system—we employ a localized spatiotemporal exclusion mechanism. This is necessary for both numerical modeling and physical processes. Spatially, an existing active storm casts an exclusion footprint (a circular zone of radius $2R_{\text{conv}}$) where the probability of new initiations is strictly zero. Temporally, once a convective event is triggered at a location, it imposes a regional ``cooldown" period on the affected grid points, locking them out of future initiations until $t_c + c_{\text{cool}}\tau_{\text{conv}}$, where $c_{\text{cool}}$ is a predefined cooldown multiplier. This is because moist convection, on both Earth and gas giants, could exhibit a recharge-discharge cycle of convective available potential energy (CAPE). The release of latent heat warms up the atmospheric interior and enhances the stratification \citep{LiIngersoll2015}; precipitation-driven downdrafts transport dry air to the lower (boundary) layer and reduce the latent heat an air parcel could release \citep{FuYang2026}.

After applying these spatiotemporal masks, a uniform random matrix is sampled against the filtered probability field to determine the exact seeding locations for the new time step, continuously updating the active event set $\mathcal{S}$. Figure~\ref{fig:convection_scheme}(b) summarizes how the smoothed vorticity determines the initiation weight, while active-event and recovery exclusions locally suppress further initiation. These local exclusions regulate event overlap and recurrence, rather than impose a mean energetic constraint. Whereas \citet{Li2006} related the mean convective forcing to radiative cooling and an observationally estimated convective area fraction, we prescribe the nominal event rate $N$ and exclusion parameters to investigate convective organization.

\subsection{The Numerical Method}
To numerically solve the continuous system, we employ a finite difference framework formulated on an Arakawa ``C" grid. Momentum advection is evaluated using a third-order upwind scheme, whose upwind bias provides grid-scale numerical dissipation while retaining higher-order accuracy. In the continuity equation, the layer-thickness fluxes are evaluated using the Quadratic Upstream Interpolation for Convective Kinematics (QUICK) scheme, which provides conservative, third-order interpolation on a uniform grid with relatively small numerical diffusion \citep{versteeg2007introduction}. For temporal integration, we utilize the second-order Adams-Bashforth (AB2) scheme to achieve a balance between accuracy and computational efficiency \citep{lomax2001fundamentals}. Because AB2 requires tendencies at both the current and preceding time levels, the first step is computed using the forward-Euler method.

To implement our convective parameterization numerically, we must discretize the convective distribution function into a probability field. For a time step $\Delta t$, the baseline discrete probability of convective initiation at grid point $\mathbf{x}$ is:

\begin{align*}
P_{\text{raw}}(\mathbf{x}, t) =\begin{cases}N \Delta t \frac{w(\mathbf{x}, t)}{\sum_{\mathbf{x}} w(\mathbf{x}, t)}, & \text{if } \sum_{\mathbf{x}} w(\mathbf{x}, t) > 0 \\\frac{N \Delta t}{N_x N_y}, & \text{otherwise}\end{cases}\end{align*}
where $N_x$ and $N_y$ are grid points in x and y directions. The uniform fallback here avoids singularity when background vorticity completely suppresses the weights.

For the discrete implementation, the exclusion condition is evaluated entirely on the Eulerian grid. A two-dimensional field $T_{\mathrm{cool},ij}$ stores the earliest time at which each grid cell may again host a convective event. At every time step, the unmasked initiation probability is first calculated from the vorticity-dependent weights. The probability is then set to zero wherever $t<T_{\mathrm{cool},ij}$. To impose spatial exclusion efficiently, the centers of all active storms are rasterized onto a binary grid, which is convolved with a circular top-hat kernel of radius $2R_{\mathrm{conv}}$; cells covered by the resulting footprint are likewise removed from the set of eligible locations. New events are sampled independently from the remaining grid-cell probabilities. Once an event is accepted, the same convolution operation identifies its surrounding exclusion footprint, within which the release time is updated to
$T_{\mathrm{cool},ij}=t_c+c_{\mathrm{cool}}\tau_{\mathrm{conv}}$.

The final initiation probability field $P^{\text{final}}(\mathbf{x}, t)$ is evaluated by masking out the currently active exclusion zones and the grid points that are still in their cooldown phase. A uniform random matrix $U(\mathbf{x}) \sim \mathcal{U}(0,1)$ is then sampled against $P^{\text{final}}(\mathbf{x}, t)$ to determine the exact seeding locations. This efficient, convolution-based process ensures physical plausibility without iterative loops.

\subsection{Parameter choice}
The initial state is set to a quiescent background with $u(\mathbf{x})=v(\mathbf{x})=0$ and $h(\mathbf{x})=H_0$. The physical domain, represented in Cartesian coordinates, covers a polar cap spanning from $70^\circ\text{N}$ to $90^\circ\text{N}$ (i.e., latitudinal extent $\theta=20^\circ$). The domain lengths $L_x$ and $L_y$ are determined by the planetary radius and the latitudinal extent, configured as $L_x = L_y =(\pi/180) a\theta/0.45 =\pi a \theta / 81$.

To identify a single convective event, we set the convective event's characteristic radius $R_{\text{conv}}$ to be proportional to the grid spacing $\Delta x$. The physical meaning in this setting is that the radius of a mature convective cell is proportional to the planetary radius. We assume the physical length scale represented by $R_{\rm conv}$, which we assume is related to nonlocal convective triggering by gravity waves, remains unconstrained. Therefore, we treat $R_{\rm conv}$ as an uncertain forcing scale and examine its influence through Group-$R_{\rm conv}$. This topic should be further discussed in the future.

Importantly, the simulation duration $T$ is extended to $40$ years. This extended time frame is specifically chosen to ensure the system reaches a steady state and to provide ample time to observe vortex drift. Furthermore, our parameter choice satisfies the CFL criterion.

\begin{deluxetable*}{llll}
\tabletypesize{\small}
\tablewidth{0pt}
\tablecaption{Reference Experimental Parameters.
\label{tab:exp_setup}}

\tablehead{
\colhead{Symbol} &
\colhead{Value} &
\colhead{Symbol} &
\colhead{Value}
}

\startdata
\cutinhead{Dimensional Parameters}
$N_x \times N_y$
& $512 \times 512$
& $c$
& $200\ \mathrm{m\,s^{-1}}$ \\
$T$
& $40\ \mathrm{yr}$
& $H_0$
& $1000\ \mathrm{m}$ \\
$a$
& $1.5 \times 10^8\ \mathrm{m}$
& $f_0$
& Varies (reference: $6.67 \times 10^{-6}\ \mathrm{s^{-1}}$) \\
$\theta$
& $20^{\circ}$
& $\nu$
& $100\ \mathrm{m^2\,s^{-1}}$ \\
$N$
& $1.39 \times 10^{-4}\ \mathrm{s^{-1}}$
& $\tau_{\rm cool}$
& $12\ \mathrm{days}$ \\
$\Delta t$
& $50\ \mathrm{s}$
& $\tau_{\rm conv}$
& $7200\ \mathrm{s}$ \\
$\tau_u=\tau_v$
& $2.09\ \mathrm{days}$
& $\tau_h$
& $2.08\ \mathrm{days}$ \\
\cutinhead{Dimensionless Parameters and Ratios}
$R_{\rm D}/L$
& $0.45$
& $c_{\rm cool}$
& $2.0$ \\
$R_{\rm conv}/\Delta x$
& $5.0$
& $c\Delta t/\Delta x$
& $0.044$ \\
$\eta$
& $5.0$
& $l_g/R_{\rm conv}$
& $2.0$ \\
$|\Phi'|/\Phi_0=|\Delta h|/H_0$
& $0.15$
& $a/L_D$
& Varies \\
\enddata

\tablecomments{
Parameters and dimensionless ratios are given for the reference
control simulation. The computational domain is square, with
$L=L_x=L_y$ and grid spacing $\Delta x=L/N_x$. $\Phi_0=g'H_0,\Phi'=g'\Delta h$.
}

\end{deluxetable*}

We use $a/L_D=Bu_p^{-1/2}$ to label the experiments and present the results, retaining a direct comparison between the planetary and deformation scales. Larger $a/L_D$ therefore corresponds to smaller $Bu_p$. The experiments are organized into six two-dimensional sensitivity groups (Table~\ref{tab:experiment_groups}). Within each group, one model parameter is varied together with $a/L_D$, while all remaining parameters are fixed at their reference values. The reference level of the group parameter therefore recovers the same five-case reference sequence across all groups. This sequence is repeated in each sensitivity figure to facilitate comparison, but the duplicated cases are counted only once, giving 75 distinct simulations
among the 100 displayed panels.

\begin{deluxetable*}{llll}

\tabletypesize{\small}
\tablewidth{0pt}

\tablecaption{
Sensitivity experiment design.
\label{tab:experiment_groups}
}

\tablehead{
\colhead{Group} &
\colhead{Motivation} &
\colhead{Varied parameter values} &
\colhead{Reference level}
}

\startdata
Group-$\Phi$
& Convective geopotential pulse amplitude
& $\Phi'\in\{-3,-6,-12\}\times10^3\ \mathrm{m^2\,s^{-2}}$
& $-6\times10^3\ \mathrm{m^2\,s^{-2}}$ \\
Group-$c_{\rm cool}$
& Temporal exclusion following convection
& $c_{\rm cool}\in\{1,2,4\}$
& $2$ \\
Group-$N$
& Expected global convective-event rate
& $N\in\{1/3600,1/7200,1/21600\}\ \mathrm{s}^{-1}$
& $1/7200\ \mathrm{s}^{-1}$ \\
Group-$\eta$
& Mesoscale feedback strength
& $\eta\in\{0,2.5,5,10\}$
& $5$ \\
Group-$l_g$
& Gaussian smoothing scale
& $l_g/R_{\rm conv}\in\{0,1,2,4\}$
& $2$ \\
Group-$R_{\rm conv}$
& Spatial scale of convective pulses
& $R_{\rm conv}/\Delta x\in\{2.5,5,10\}$
& $5$
\enddata

\tablecomments{
Each group is a two-dimensional sweep in which the parameter listed in
the third column is varied jointly with
$a/L_D\in\{1,2.5,5,10,20\}$, while all remaining parameters retain their
reference values. The five cases in which all parameters other than
$a/L_D$ take their reference values form a common reference sequence
shared by all six groups. Its $a/L_D=5$ member is the nominal reference
case. Consequently, the 100 panels displayed across the six sensitivity
figures correspond to 75 distinct simulations.
}

\end{deluxetable*}

\section{Results}\label{Section:3}
\subsection{Vortex size}
The 75 simulations give 75 final-state fields and their evolution. We use Fourier transforms to analyze each field (see Appendix~\ref{Appendix B1}). The kinetic energy spectrum is obtained by summing the squared Fourier amplitudes of both velocity components over annular wavenumber shells, after removing the axisymmetric polar circulation. The wavenumber corresponding to the spectrum's peak, $k_{\rm E,peak}$, is used to estimate a characteristic vortex radius, $R_{\rm E}=\pi/k_{\rm E,peak}$. This half-wavelength convention provides a common measure across all simulations without identifying individual vortex centers. Where distinct vortices are absent, it should instead be interpreted as a characteristic flow scale. To investigate the scale selection mechanism, we perform a linear stability analysis.

%consider the linear instability of the averaged convective-feedback model.

For small, balanced perturbations proportional to $\exp(i\mathbf K\boldsymbol{\cdot}\mathbf x+\gamma t)$, where $K=|\mathbf K|$ is the horizontal angular wavenumber and $\gamma$ is the growth rate, the analysis in Appendix~\ref{Appendix A} gives
\begin{equation}
\gamma(K)=\frac{1}{\tau_{\rm cool}}\frac{\Pi\eta L_D^2K^2\exp(-K^2\ell^2/4)-1}{1+L_D^2K^2},\qquad \ell^2=l_g^2+2R_{\rm conv}^2.
\label{eq:results_linear_growth}
\end{equation}
Here, $\Pi=-\tau_{\rm cool}\overline Q_{\rm conv}/H_0$ measures the mean convective mass sink relative to radiative relaxation, and $\ell$ combines the Gaussian smoothing scale with the spatial scale of convective pulses. This early-growth approximation neglects nonlinear advection, the spatial variation of $f$, and boundary and viscous effects.

The model admits both stable and unstable responses. When $\Pi\eta\leq e\ell^2/(4L_D^2)$, there is no positive-growth band: the smoothed convective feedback cannot overcome radiative relaxation at any wavenumber. An unstable band exists when
\begin{equation}
\Pi\eta>\frac{e\ell^2}{4L_D^2}.
\label{eq:results_instability_threshold}
\end{equation}
Thus, convection does not necessarily produce a self-amplifying organized disturbance. In particular, $\eta=0$ admits no growing mode in this model, although stochastic convection can still generate transient motions in the simulations.

\begin{figure*}[t]
\centering
\includegraphics[width=0.85\textwidth]{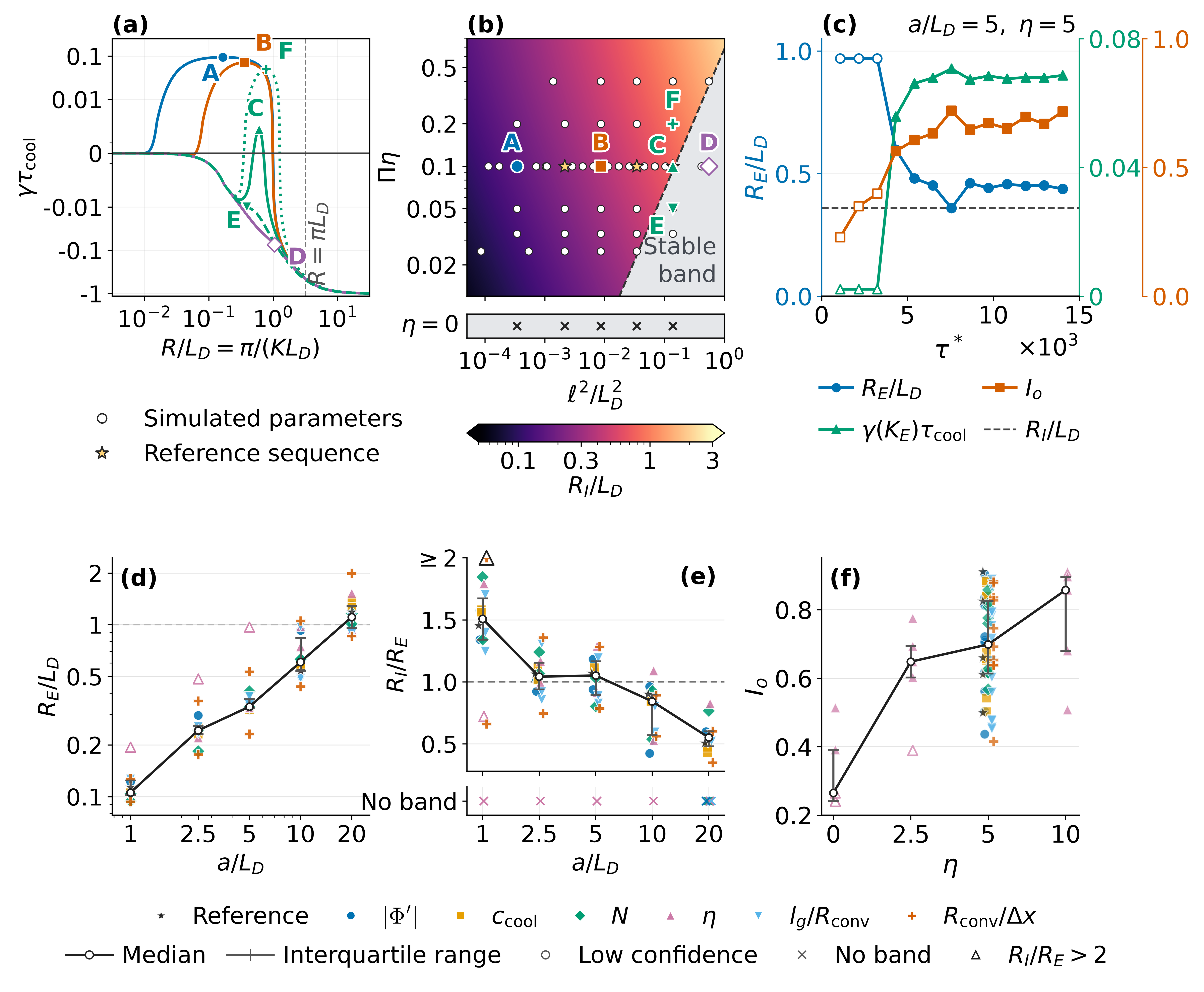}
\caption{Scale selection and spectral organization. (a) Linear growth curves; symbols identify curves, marking positive maxima where present. (b) Approximate $R_I/L_D$; gray denotes no unstable band, circles simulated parameters, and stars the reference sequence. D is an illustrative, unsimulated combination. (c) Early evolution of the reference case: $R_E/L_D$ uses the blue left axis, $\gamma(K_E)\tau_{\rm cool}$ uses the green inner-right axis, and the organization index $I_o$ (to be introduced in section \ref{3.2}) uses the orange outer-right axis; the dashed line marks $R_I/L_D$. The 13 snapshots span days 90--1170; no values are assigned before the first saved field. The nondimensional time is $\tau^*=t/\tau_{\rm conv}$, with $\tau_{\rm conv}=7200\,\mathrm{s}$. (d--f) Mature $R_E/L_D$, $R_I/R_E$, and $I_o$. Case-specific $R_I$ values use numerical growth-rate maxima. Black curves and bars show medians and interquartile ranges; open symbols mark low-confidence peaks. Crosses denote no unstable band, and ratios above 2 are capped in (e). Colors and symbols distinguish experimental groups.}
\label{fig:vortex_scale_organization}
\end{figure*}

Figure~\ref{fig:vortex_scale_organization}(a) illustrates both responses, with the normalized half-wavelength radius $R/L_D=\pi/(KL_D)$ increasing from left to right. The condition $KL_D=1$ corresponds to $R/L_D=\pi$, not unity. Matching letters in panels (a,b) link each growth curve to its parameters: A--D hold $\Pi\eta=0.09986$ fixed, while E and F halve and double this product relative to C at the same smoothing ratio. The ordinate in (a) is symmetric logarithmic, with a linear interval around zero. For a fixed convective forcing, feedback parameter, and cooling time scale, disturbances much smaller than $\ell$ receive little coherent convective amplification because their vorticity anomalies are smoothed out in the feedback. As the disturbance scale increases beyond the smoothing scale, this attenuation weakens, allowing instability. In the presence of scale separation ($\ell\ll L_D$), this unstable range lies between the smoothing and deformation scales.

At still larger scales, the partition of the potential-vorticity anomaly between relative vorticity and thickness becomes important. Under the quasigeostrophic approximation,
\begin{equation*}
\widehat q'=\frac{\widehat\zeta'}{H_0}-\frac{f_0\widehat h'}{H_0^2},\qquad
\frac{\widehat\zeta'}{H_0}=\frac{L_D^2K^2}{1+L_D^2K^2}\widehat q',\qquad
-\frac{f_0\widehat h'}{H_0^2}=\frac{1}{1+L_D^2K^2}\widehat q'.
\end{equation*}
Equivalently, \(L_D^2K^2\propto-\widehat\zeta'/\widehat h'\) for a fixed background state. Thus, when the disturbance scale is smaller than \(L_D\), most of its potential-vorticity anomaly is expressed through relative vorticity. As the scale approaches and exceeds \(L_D\), potential vorticity gets dominated by thickness displacement rather than relative vorticity. The resulting weak vorticity signal cannot sustain the vorticity-dependent convective feedback. Meanwhile, the dominance of thickness displacement indicates that radiative relaxation eventually dominates, and \(\gamma\rightarrow-1/\tau_{\rm cool}\) in the long-wave limit. 

%The transition need not occur exactly at \(L_D\), because weak feedback can stabilize disturbances that remain smaller than \(L_D\). These curves compare disturbances of different scales rather than describe the temporal growth of an individual vortex.

The product $\Pi\eta$ determines how fast this intermediate growing range develops. A convective mass sink produces convergence and increases cyclonic vorticity, which increases the probability of further nearby convection. Strengthening this feedback can turn decay into growth, broaden the unstable band, and shift its maximum, as illustrated by curves E, C, and F. If $\Pi\eta$ is held fixed, changing $\tau_{\rm cool}$ only rescales the growth rates and leaves the preferred wavenumber unchanged. The simulations considered here keep $\tau_{\rm cool}$ fixed.

Whenever an unstable band exists, it contains a most unstable wavenumber, $k_{\rm I,peak}$. We define the corresponding radius as half of the most unstable wavelength, $R_{\rm I}=\pi/k_{\rm I,peak}$. Under the approximations $k_{\rm I,peak}\ell\ll1$ and $\ell/L_D\ll1$ described in Appendix~\ref{Appendix A},
\begin{equation}
\frac{R_{\rm I}}{L_D}\simeq\pi\left[\sqrt{1+\frac{4L_D^2}{\ell^2}\left(1+\frac{1}{\Pi\eta}\right)}-1\right]^{-1/2}.
\label{eq:results_linear_radius}
\end{equation}
Panel (b) shows this approximate scale across parameter space, together with the instability boundary. Within the asymptotic regime, $R_{\rm I}\propto(\ell L_D)^{1/2}[1+1/(\Pi\eta)]^{-1/4}$: the preferred radius depends jointly on the deformation radius, smoothing, and feedback strength. A similar geometric-mean scale was obtained for tropical cyclone embryos using moisture-dependent heating with nonlocal cold-pool effects \citep{Fu2026ColdPools}; while both models recover an $f^{-1/2}$ dependence (equivalently, an $L_D^{1/2}$ dependence at fixed gravity-wave speed $c$) in the appropriate asymptotic regime, our vorticity-dependent feedback and Newtonian cooling introduce an additional dependence of the preferred scale on feedback strength. Using vorticity rather than moisture as the feedback variable also avoids specifying how the abundances of individual condensable species regulate convection in the present idealized Jovian application, with these species-dependent effects left for future investigation. At fixed forcing parameters, decreasing $L_D$ reduces $R_{\rm I}$ more slowly than $L_D$ itself, so $R_{\rm I}/L_D$ increases. Increasing the effective smoothing scale $R_{\rm conv}$ favors larger disturbances, provided instability persists. Stronger feedback likewise increases the preferred radius toward a limiting value in this approximation. For the case-by-case comparisons, we search for the most unstable growth rate numerically, rather than using Equation~\ref{eq:results_linear_radius}.

Panel (c) follows the earliest saved development of the reference case, comparing the simulation ($R_E/L_D$) with the theory ($\gamma(K_E)\tau_{\rm cool}$), where $K_E\equiv k_{\rm E,peak}=\pi/R_E$ and $\gamma(K_E)$ is the theoretical growth rate evaluated at the evolving spectral peak. The spectral scale approaches $R_I$ as the spectrum becomes more concentrated; however, the earliest peaks lie at the diagnostic band edge, and the closest agreement occurs after nonlinear effects become appreciable, so this evolution provides qualitative support for scale selection rather than a direct validation of the growth rates.

Panels (d,e) compare these predictions with the mature energy-spectrum scale. The median $R_{\rm E}/L_D$ increases from approximately $0.11$ at $a/L_D=1$ to $1.11$ at $a/L_D=20$, showing that the mature vortex scale does not follow a simple proportionality to $L_D$. The most unstable and mature radii are comparable at $a/L_D\simeq2.5$--$5$, where the median $R_{\rm I}/R_{\rm E}$ is about $1.05$. At smaller $a/L_D$, the mature vortex scale is generally smaller than $R_{\rm I}$; the median ratio is $1.51$ at $a/L_D=1$. The opposite tendency occurs at larger $a/L_D$, with median ratios of $0.84$ and $0.55$ at $a/L_D=10$ and $20$.

The linear stability analysis therefore provides a useful reference for the mature vortex scale. Fragmentation of large vortices at low $a/L_D$ can reduce the mature vortex scale relative to the initial growth scale, whereas the nonlinear cascade of kinetic energy toward larger scales may make $R_{\rm E}>R_{\rm I}$ at larger $a/L_D$. These possible explanations concern subsequent nonlinear evolution, which the linear analysis does not describe.

%, particularly in the intermediate regime

\subsection{Vortex structure}\label{3.2}
\begin{figure*}[t]
\centering
\begin{interactive}{animation}{final_pv_eta_4k.mp4}
\includegraphics[width=\hsize]{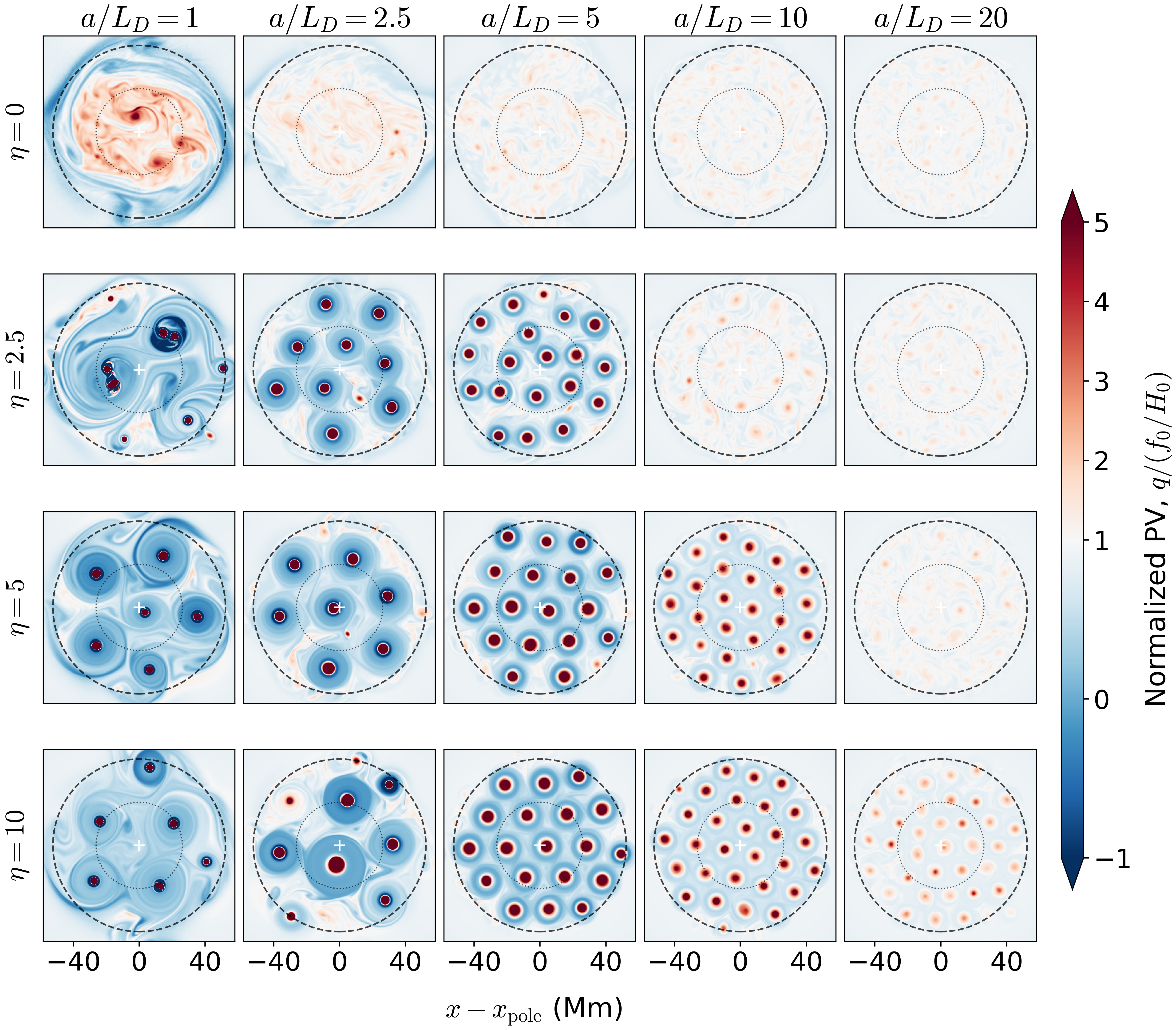}
\end{interactive}
\csname digitassetfalse\endcsname
\caption{Final potential-vorticity fields at 40 years for the mesoscale-feedback experiments. Columns vary $a/L_D$ and rows vary $\eta$; other parameters take their reference values. PV is normalized by $f_0/H_0$, with unity shown in white. The outer dashed and inner dotted circles mark effective latitudes of $70^\circ\mathrm N$ and $80^\circ\mathrm N$, respectively; the outer circle is also the physical-domain boundary. Effective latitude is defined by $\phi_{\rm eff}=90^\circ-(180^\circ/\pi)(r/a)$ as a spherical reference for the planar domain. Coordinates are pole-centered, and $1\,{\rm Mm}=10^6\,{\rm m}$. The complete Figure Set (6 images), showing the corresponding final fields for all 75 simulations, is available in the online journal. An animated version of this figure is also available, showing the evolution from approximately 0.25 to 40 model years with the same panel arrangement and color scale.}
\label{fig:final_field}
\end{figure*}

The corresponding final-state maps for all 75 simulations are provided in the online Figure Set. Figure~\ref{fig:final_field} shows results of Group-$\eta$, and the regime transition with deformation radius and mesoscale feedback factor. Without mesoscale feedback, vortex growth is slow and irregular, and a coherent ``shield'' is often absent. Increasing the feedback strength accelerates vortex organization and strengthens the surrounding low-PV shield. The separation between compact high-PV cores and broader low-PV surroundings is particularly clear at $a/L_D=2.5$ and $5$, where several cases form persistent vortex crystals. The shielded structure helps the vortices remain distinct rather than merge directly, consistent with the stabilizing role of shields discussed by \citet{Li2020}.

The time evolution provides further information beyond the final fields. When $a/L_D$ is sufficiently small, a growing vortex can span much of the domain before its contracting core becomes unstable and fragments into several smaller vortices. As $a/L_D$ increases, the characteristic vortex size generally decreases, allowing the domain to accommodate more vortices where organization is sustained. At $a/L_D=20$, weak-feedback cases remain poorly organized, whereas stronger feedback supports more distinct cores. This trend is consistent with the slower growth near marginal instability in the linear model, although that model does not describe mature-core instability or shield formation.

\citet{CHEN2025116438} found that a sufficiently large $L_D$ favors shielded vortex crystals in decaying shallow-water turbulence by limiting relative layer-thickness variations. In our continuously forced simulations, organization also depends on convective feedback, as shown by the different patterns obtained at fixed $a/L_D$ when $\eta$ varies. Because the forcing and parameter variations differ, the crystal-forming parameter ranges are not directly comparable between the two models.

%Thus, the $a/L_D=1$ experiments do not simply approach a persistent Saturn-like single-vortex state. 

%higher feedback threshold and

Further increases in $\eta$ do not necessarily produce a proportionate increase in shield strength. The total expected convective event is bounded in the numerical model through the exclusion principle (see Fig. \ref{fig:convection_scheme}), while smoothing, finite event size, and spatiotemporal exclusion limit the localization and recurrence of convection. These constraints provide a possible explanation for the weaker visual changes between some of the high-feedback cases.

To complement these morphological differences, we quantify the concentration of kinetic energy around the spectral peak using the organization index
\begin{equation}
I_o=\left[1+\sum_{j\in\mathcal B}p_j\left(\ln\frac{k_j}{k_{\rm E,peak}}\right)^2\right]^{-1},\qquad p_j=\frac{E_j}{\sum_{j\in\mathcal B}E_j},
\label{eq:results_organization}
\end{equation}
where $E_j$ is the energy in the shell centered at $k_j$, $p_j$ is its fraction of the energy in the diagnostic band $\mathcal B$, and $k_{\rm E,peak}$ is the peak wavenumber. The band and implementation are specified in Appendix~\ref{Appendix B1}. A narrow spectrum gives $I_o$ close to unity, while energy spread over different relative scales lowers it. Thus, $I_o$ measures spectral concentration rather than directly measuring shield strength or spatial lattice order.

Figure~\ref{fig:vortex_scale_organization}(f) shows that higher $\eta$ is associated with higher $I_o$, with the median increasing from $0.27$ without feedback to $0.86$ at $\eta=10$. Increasing $\eta$ concentrates more of the fixed global event supply within cyclonic regions, favoring persistent vortex growth over scattered convection. When vortices develop comparable characteristic sizes, their energy is concentrated around a common wavenumber. The spread at fixed $\eta$ reflects the influence of the other parameters and subsequent fragmentation, deformation, and vortex interactions; the index alone cannot separate these effects. The samples at $\eta=0$, $2.5$, $5$, and $10$ contain 5, 5, 60, and 5 cases, respectively; their medians therefore summarize unequal samples rather than a balanced comparison across all parameters.

With sufficiently effective convective organization, the vortices can occupy much of the physical domain and remain separated by their shields. Their subsequent collective motion is considered in the following subsection.

\subsection{Structural  drift}
An animated version of Figure~\ref{fig:final_field} shows the evolution of the $\eta$ experiments, including vortex formation, reorganization, and structural drift. When the vortices do not fill the domain, they exhibit repeated close encounters, making their patterns unstable. For the cases that form statistically steady vortex crystals, the patterns undergo structural drift. A primary idea of such drift is the beta-gyre, which is produced by the vortex's advection of the meridional potential vorticity gradient. It makes the vortex drift northwestward in the Northern Hemisphere \citep{gavriel2023}. Essentially, the beta-gyre is produced by Rossby waves; thus, the phase speed of Rossby waves is a reasonable reference for the drift angular velocity. We can express it as
\begin{equation*}
    c_R=-\frac{\beta}{k^2+L_D^{-2}},
\end{equation*}
where $k\sim R_m^{-1}$ is estimated using the radius of maximum tangential velocity $R_m$, obtained from the fitted vortex velocity profile (see Appendix~\ref{Appendix B2}). Substitute $\beta=f_0r/a^2$ and we get
\begin{equation*}
    \omega_R=\frac{c_R}{r}=-\frac{f_0}{a^2(R_m^{-2}+L_D^{-2})}.
\end{equation*}

The structural drift angular velocity is diagnosed from the evolution of the entire vortex pattern using FFT-based azimuthal cross-correlation between adjacent and second-neighbor frames (see Appendix~\ref{Appendix B3}). The result is shown in Table~\ref{tab:structural_rossby_comparison}. The Rossby wave speed estimate is clockwise in every case, whereas the median structural drift is counterclockwise for $a/L_D=1$ and $2.5$. Even when the directions match, the structural rotation is generally several times faster than the Rossby wave speed estimate. Across all 39 reliable cases, the median $|\omega_{\rm R}/\omega_{\rm str}|$ is $0.084$, and the correlation between the two angular velocities remains weak ($r=0.210$). Therefore, a conventional beta-drift scale cannot by itself account for either the direction or magnitude of the diagnosed structural drift. It is also notable that Rossby wave speed is deduced under quasi-geostrophic assumption, which may not be true in some cases where Rossby number $\rm{Ro}$ is much larger than 1.

\begin{table}[htbp]
\centering
\renewcommand{\arraystretch}{1.15}
\begin{tabular}{ccccc}
\hline
$a/L_D$ &
$N$ &
CW/CCW/S &
$\omega_{\rm str}$ &
$\omega_{\rm R}$ \\
 & & &
\multicolumn{2}{c}{(${\rm deg\,yr^{-1}}$)} \\
\hline
$1.0$  & 6  & 1/5/0 & $ 62.1\ [20.7,\ 103.5]$  & $-0.31\ [-0.40,\ -0.25]$ \\
$2.5$  & 10 & 4/6/0 & $ 17.2\ [-36.8,\ 44.6]$  & $-1.39\ [-2.30,\ -1.04]$ \\
$5.0$  & 12 & 9/2/1 & $-22.2\ [-38.3,\ -8.9]$  & $-3.70\ [-5.20,\ -2.76]$ \\
$10.0$ & 10 & 9/1/0 & $-37.1\ [-41.5,\ -19.7]$ & $-8.72\ [-8.94,\ -5.81]$ \\
$20.0$ & 1  & 1/0/0 & $-24.0$                    & $-14.66$ \\
\hline
\end{tabular}
\caption{Comparison between the diagnosed structural angular velocity $\omega_{\rm str}$ and the Rossby wave phase-speed estimate $\omega_{\rm R}$. Values are medians with interquartile ranges in brackets over cases with reliable structural-drift measurements. CW, CCW, and S denote clockwise, counterclockwise, and stationary patterns, respectively; positive angular velocity is counterclockwise. The $a/L_D=20$ result contains only one reliable case and is therefore not statistically representative.}
\label{tab:structural_rossby_comparison}
\end{table}

\citet{BENZEGGOUTA2026119877} pointed out that vortex drift is affected not only by beta gyres but also by mutual advection, which motivates us to use the relative-vorticity field to estimate the velocity induced across each positive-vorticity core. For a source region $\mathcal R$, the induced velocity field and its core-vorticity-weighted mean velocity are
\begin{align}
\mathbf u^{(\mathcal R)}(\mathbf x)&=\frac{1}{2\pi}\int_{\mathcal R}\zeta(\mathbf x')\frac{\hat{\mathbf z}\times(\mathbf x-\mathbf x')}{|\mathbf x-\mathbf x'|^2}\,\mathrm dA',\\
\mathbf U_i^{(\mathcal R)}&=\frac{1}{\Gamma_i}\int_{\mathcal C_i}\zeta(\mathbf x)\mathbf u^{(\mathcal R)}(\mathbf x)\,\mathrm dA,
\qquad \Gamma_i=\int_{\mathcal C_i}\zeta(\mathbf x)\,\mathrm dA,
\end{align}
where $\mathcal C_i$ is the positive-vorticity core of vortex $i$, $\mathbf x'$ and $\mathbf x$ are the source and evaluation positions, respectively, and $\hat{\mathbf z}$ is normal to the model plane.

To further diagnose the origin of the structural drift, each vortex is represented by an object $\mathcal O_i=\mathcal C_i\cup\mathcal S_i$, where $\mathcal S_i$ is its accepted anticyclonic shield. We then divide the source field into the target object itself, the union of the other vortex objects, and the residual background outside all objects. If no shield boundary can be identified reliably, the corresponding object contains only its positive-vorticity core. The velocity is averaged over the target core rather than sampled only at its center. Under this weighting, the core's self-induced contribution cancels by antisymmetry, so the nonzero self-object term measures the steering of the core by its own asymmetric shield. The corresponding pattern angular velocities are obtained from the least-squares rigid-rotation component (see Appendix~\ref{Appendix B4}).

This diagnosis only uses the realized relative-vorticity field and should therefore be interpreted as a spatial decomposition of rotational advection, rather than as a decomposition into independent forcing mechanisms. But the advection includes many physical processes. For example, beta gyres arise from advection of the planetary vorticity gradient and evolve through interaction with the vortex circulation, producing an asymmetric flow that steers the vortex \citep{LiWang1994}. Their induced velocity is included in the self-object term when the resulting anomaly lies within the accepted shield, and in the residual background when it lies outside the identified vortex objects; adding a separate beta-drift velocity would therefore risk double counting. The other-object term represents mutual advection by the coherent vortices, whereas the residual contains interstitial vorticity, filaments, unresolved turbulence, and the large-scale circulation. We additionally split the residual by the nominal disk boundary to audit the direct contribution of exterior vorticity. Damping where a vortex overlaps the sponge may nevertheless modify the vortex object or the interior residual and is therefore not represented by the exterior contribution alone.

\begin{figure*}[t]
\centering
\includegraphics[width=0.88\textwidth]{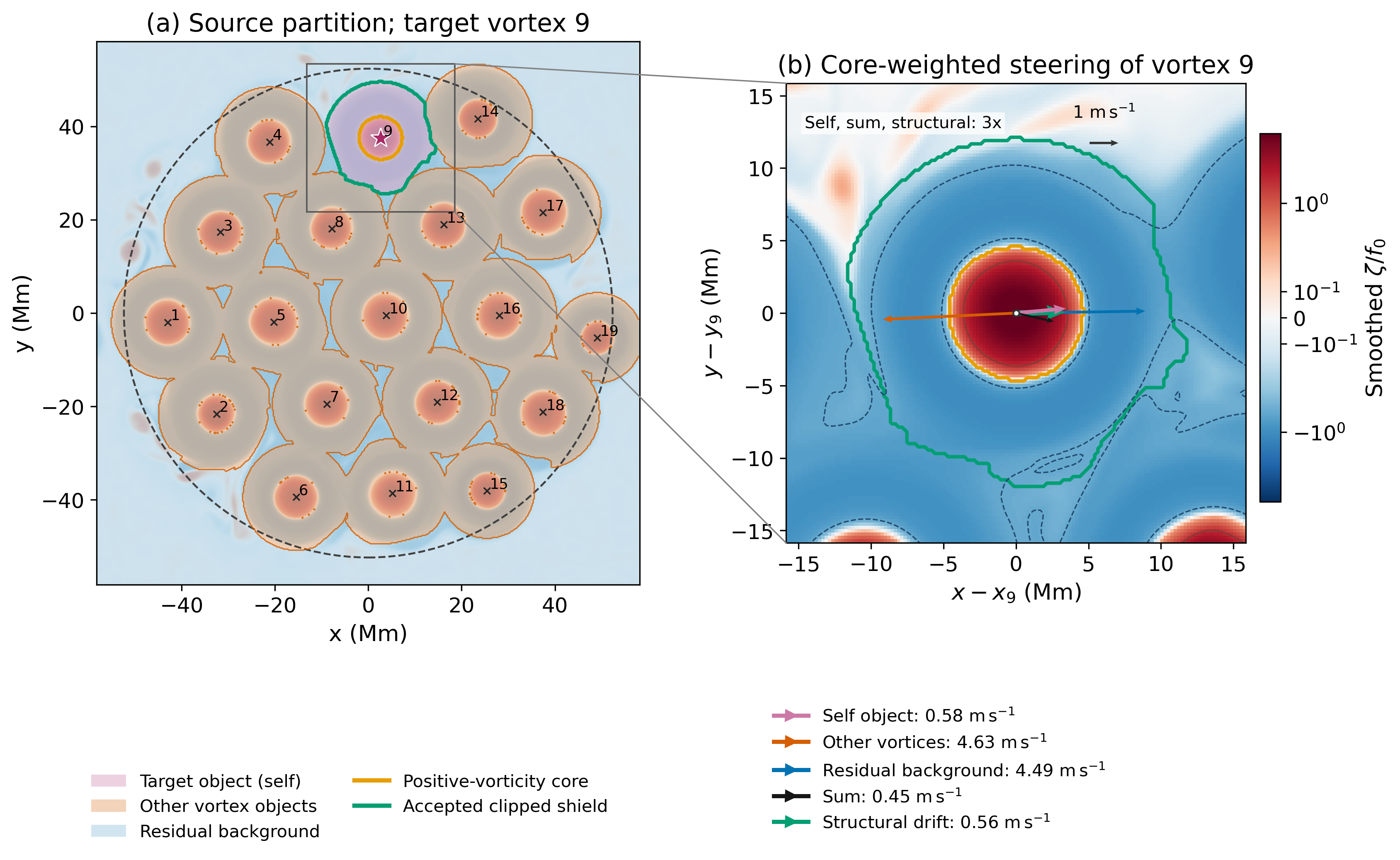}
\caption{Object-based Biot--Savart diagnosis for vortex 9 in the final state of the $a/L_D=5$, $\eta=10$ experiment; all other parameters take their reference values. (a) Explicit source partition for target vortex 9. Magenta denotes the target object $\mathcal O_9=\mathcal C_9\cup\mathcal S_9$, orange denotes the union of the other vortex objects, and blue denotes the residual background outside all identified objects. The orange and green curves delimit the target's positive-vorticity core $\mathcal C_9$ and accepted shield $\mathcal S_9$, respectively. The dashed circle marks the nominal disk; objects may extend across it, and it separates only the interior and exterior audit portions of the residual. All 19 centers are numbered, and the box locates panel (b). The translucent partition is overlaid on the one-grid-cell-smoothed $\zeta/f_0$ field. (b) Core-vorticity-weighted velocities induced by the three source classes, together with their sum and the FFT structural-drift reference. Before plotting, the mean of each contribution over all centers is removed to suppress bulk translation. The self-core contribution cancels under the adopted weighting, so the self-object arrow is the contribution of the target's accepted shield. Thin contours mark $\zeta/f_0=-0.5$ and $2$. The structural reference uses $\omega_{\rm str}=-26.57\ \mathrm{deg\,yr^{-1}}$ over years 30.08--40, whereas the induced velocities use the year-40 field. The other-vortex and residual-background arrows follow the displayed $1\ \mathrm{m\,s^{-1}}$ scale; the self-object, summed, and structural-drift arrows are enlarged together by a factor of three for visibility. Their physical magnitudes are listed in the legend. Coordinates are relative to the pole in (a) and to vortex 9 in (b); $1\ \mathrm{Mm}=10^6\ \mathrm{m}$.}
\label{fig:biot_savart_eta10_vortex9_example}
\end{figure*}

As an example, Figure~\ref{fig:biot_savart_eta10_vortex9_example} illustrates the decomposition for the $a/L_D=5$, $\eta=10$ experiment. Panel (a) explicitly maps the three source classes relative to target vortex 9: the target object itself, the other vortex objects, and the residual background. The nominal-disk boundary further separates the interior and exterior portions of the residual without treating them as independent additional regions. Panel (b) shows the corresponding core-weighted induced velocities. For vortex 9, the self-object, other-vortex, and residual-background speeds are $0.580$, $4.631$, and $4.487\ \mathrm{m\,s^{-1}}$, respectively. The two large environmental terms nearly cancel, leaving a summed speed of $0.449\ \mathrm{m\,s^{-1}}$, compared with the structural-rotation reference of $0.565\ \mathrm{m\,s^{-1}}$, with a directional difference of $13.67^\circ$. For the complete 19-vortex pattern, the induced angular velocity is $-13.45\ \mathrm{deg\,yr^{-1}}$, compared with $-26.57\ \mathrm{deg\,yr^{-1}}$ from the FFT diagnostic. Thus, this example recovers the clockwise direction but underestimates the rate, while showing explicitly how a small net drift can arise from cancellation between much larger induced velocities.

The diagnostic result reveals a competition between the source classes. Among the 38 reliable nonstationary cases, the self-object term is clockwise in 30 cases, counterclockwise in 7, and zero in 1, with a median absolute magnitude of $15.7\ \mathrm{deg\,yr^{-1}}$. In contrast, the other-vortex term is counterclockwise in all 38 cases and the residual background is clockwise in all 38, with median absolute magnitudes of $228.0$ and $196.1\ \mathrm{deg\,yr^{-1}}$, respectively. The residual is dominated by its portion inside the nominal disk (median absolute magnitude $196.4\ \mathrm{deg\,yr^{-1}}$); the exterior residual portion is much smaller (median $1.8\ \mathrm{deg\,yr^{-1}}$). Structural drift is therefore the residual of strong collective advection by the other coherent vortices and an opposing contribution from the interstitial background, with a usually smaller correction from the target's own shield. The total induced rotation reproduces the observed direction in $35/38=92.1\%$ of the reliable nonstationary cases, correlates with the diagnosed structural angular velocity ($r=0.82$), and has a median magnitude ratio $|\omega_{\rm BS}/\omega_{\rm str}|=0.92$. Because the large other-vortex and residual terms nearly cancel, their individual amplitudes are more sensitive to the object boundaries than their sum. Details and the population comparison are given in Appendix~\ref{Appendix B4}. The direct velocity induced by exterior residual vorticity is secondary, but this does not exclude indirect boundary effects. Moreover, the free-space relative-vorticity diagnostic omits divergent flow and does not identify the tendency processes that create the diagnosed asymmetries. The drift velocities in this model therefore remain only partially explained, and a fuller dynamical attribution is left for future work.

\section{Conclusion}\label{Section:4}
In summary, we investigated polar vortex formation using a 1.5-layer shallow-water model on a polar $\gamma$ plane, adapting the vorticity-dependent convective feedback of \citet{FuONeill2025c}. Starting from a quiescent state, localized convective mass sinks generate cyclones with surrounding low-PV shields that can organize into persistent vortex crystals. Previous studies examined polar vortex generation under stochastic forcing \citep{Brueshaber2019} and demonstrated the importance of shielding for the stability of initialized vortex arrays \citep{Li2020}. Our simulations connect these aspects by providing a possible formation pathway for shielded vortex crystals through feedback between convection and the evolving circulation.

The sensitivity experiments confirm the importance of the planetary Burger number for polar vortex configurations, consistent with the planetary Burger number dependence identified by \citet{Brueshaber2019}. They also show that mesoscale feedback promotes vortex growth, shield formation, and sharpens the energy spectrum. The characteristic radius of the mature vortex is not a fixed multiple of the deformation radius, indicating the influence of additional scales associated with convection. In the $a/L_D\simeq2.5$--$5$ regime, the mature vortices' scale diagnosed from the Fourier spectrum is comparable to the most unstable radius of the linear stability analysis, providing a useful reference for scale selection while leaving room for subsequent nonlinear evolution. The linear stability analysis also reveals that the most unstable wave number depends on the deformation radius, the convective feedback's smoothing length, and the feedback strength. However, linear stability analysis is used only to study the scale and motion of vortices, rather than to predict vortex strength, because intense vortices exhibit strong nonlinearity.

The simulated vortex crystals exhibit structural drift in both directions, whose magnitude and direction are not adequately represented by the adopted Rossby wave phase-speed estimate. The object-based Biot--Savart diagnostic finds a usually modest clockwise contribution from a vortex's own shield, a strong counterclockwise contribution from the other coherent vortex objects, and an opposing clockwise contribution from the residual background. Their substantial cancellation controls the net drift. The total induced rotation reproduces the measured direction in $92.1\%$ of the reliable nonstationary cases, correlates with the diagnosed structural angular velocity ($r=0.82$), and has a median magnitude ratio of $0.92$. The small direct contribution from exterior residual vorticity suggests that boundary-region vorticity is not the leading source of the induced velocity, although indirect boundary effects remain possible. The importance of mutual advection is qualitatively consistent with the laboratory results of \citet{BENZEGGOUTA2026119877}, in which mutual advection can produce eastward crystal drift.

Observations provide a quantitative reference for the simulated structural drift. JunoCam tracking during the first two years of the mission revealed a systematic westward drift of the southern circumpolar cyclones at $1.5\pm0.2^\circ$ per 53-day orbit, equivalent to approximately $10.3\pm1.4^\circ\,\mathrm{yr^{-1}}$, whereas no consistent longitudinal drift was identified in the northern cyclones over that interval \citep{TabatabaVakili2020}. Longer-term JIRAM observations indicate mean westward rates of approximately $3^\circ\,\mathrm{yr^{-1}}$ and $7^\circ\,\mathrm{yr^{-1}}$ at the north and south poles, respectively \citep{Mura2022,gavriel2023}. In comparison, the reliable simulated cases at $a/L_D=5$ and $10$ have median structural angular velocities of $-22.2^\circ\,\mathrm{yr^{-1}}$ and $-37.1^\circ\,\mathrm{yr^{-1}}$, respectively (Table~\ref{tab:structural_rossby_comparison}). Their westward direction agrees with the observed long-term tendency, but their median rates are appreciably faster than these observational estimates. At $a/L_D=1$ and $2.5$, the simulated medians are instead eastward, at $62.1^\circ\,\mathrm{yr^{-1}}$ and $17.2^\circ\,\mathrm{yr^{-1}}$. Thus, the simulations produce persistent crystals with westward drift, but do not robustly reproduce the slow westward preference observed on Jupiter across the explored parameter range.

These differences must nevertheless be interpreted with care. \citet{gavriel2023} explicitly analyzed the motion of the vortex crystal's center of mass and related its westward precession to the observed cyclone drift. Our FFT diagnostic instead measures changes in pattern orientation after recentering, which is not identical to either the individual longitudinal drift rates or the center-of-mass trajectory. Recentering removes bulk translation but does not, by itself, separate the dynamical mechanisms responsible for the remaining rotation. The observational rates therefore provide a useful benchmark, while a like-for-like comparison requires applying a common rotation diagnostic to the observed and simulated configurations. Within the simulations, the adopted Rossby phase-speed estimate does not capture the range of drift directions and rates, while the Biot--Savart diagnostic identifies strong, opposing contributions from the other vortices and the residual background. This suggests that collective advection must be considered when interpreting crystal rotation, without excluding $\beta$-gyres, whose induced velocities are already represented in the realized vorticity field. A corresponding separation of pattern rotation and collective displacement in JunoCam and JIRAM observations would allow a more direct comparison. Whether the same competition produces Jupiter's slower, persistently westward drift remains to be established.

An important extension is to examine interactions between the polar vortex system and the surrounding subpolar jet system. Observations have revealed turbulent energy transfers at Jovian high latitudes and characterized the surrounding folded-filamentary regions \citep{Siegelman2022-2,Fletcher2026}, but further observations and modeling are needed to establish their influence on crystal stability and structural drift. A parameterization of turbulence at the domain's lateral boundary might represent the averaged influence of these surroundings, although the present sponge provides only prescribed damping and its suitability for this purpose remains untested. Varying the sponge location, width, and damping rates, followed by simulations with dynamically active surroundings in a larger spherical domain, would help assess this possibility. The small direct contribution from exterior residual vorticity does not exclude an indirect influence through modifications of the interior circulation.

The applicability of the convective parameterization also requires further assessment. Its prescribed dependence on vorticity, convective event radius, and recurrence times should be tested against observations and simulations that explicitly resolve moist convection and vertical coupling. Such comparisons would help establish which aspects of shield formation and spectral scale selection persist across different convective closures; the crystal-forming parameter ranges obtained here therefore remain model-dependent rather than direct constraints on Jupiter's deformation radius. Vortex interactions provide another direction for investigation. Controlled experiments varying vortex spacing and shielding, combined with vorticity-tendency budgets, could clarify how shield deformation, beta-gyres, and mutual advection establish persistent spacing or lead to encounters, rearrangements, and changes in structural drift. These experiments would extend the present object-based spatial decomposition of rotational advection toward identifying the processes that generate the diagnosed asymmetries.

\section*{Data Availability Statement}
The simulation code, final-state model fields, for all 75 simulations, and figures and animations of all six experimental groups are publicly available in the Zenodo repository (\href{https://doi.org/10.5281/zenodo.23632380}{doi: 10.5281/zenodo.23632381}). The complete Figure Set and an animation of the Group - $\eta$ experiments accompanying Figure~\ref{fig:final_field} are also available in the online journal.

\begin{acknowledgments}
Yuhang Ruan is supported by the College Students’ Innovative Training Plan of Nanjing University, project number 202610284180. Hao Fu is supported by the faculty startup funding from Nanjing University. Cheng Li is supported by Heising-Simons Foundation. Da Yang is supported by the Packard Fellowship for Science and Engineering and an NSF CAREER Award (AGS-2048268). The computational resources used in this work were provided by the High Performance Computing Center of Nanjing University. The authors used OpenAI’s ChatGPT and Codex, with the GPT-5.6 Sol, GPT-6.0 Sol and GPT-6 Astra models, during manuscript preparation to assist with language editing, organization of the presentation, generation and refinement of code, and data visualization. Figure~\ref{fig:convection_scheme} was generated using OpenAI’s image-generation tools through ChatGPT under the supervision of the authors. All scientific arguments, methodological choices, numerical results, interpretations, and final text were reviewed and verified by the authors.
\end{acknowledgments}

\appendix
\section{Linear Instability of the Convective Feedback}\label{Appendix A}

We examine whether the vorticity-dependent convective feedback can select a preferred horizontal scale during the early development of the flow. We consider the shallow-water equations without the sponge, dissipation, or planetary-vorticity gradient:
\begin{align}
\frac{\partial\mathbf{V}}{\partial t}+(\mathbf{V}\boldsymbol{\cdot}\nabla)\mathbf{V}+f_0\hat{\mathbf{k}}\times\mathbf{V}&=-g'\nabla h,
\label{eq:linear_sw_momentum}\\
\frac{\partial h}{\partial t}+\nabla\boldsymbol{\cdot}(h\mathbf{V})&=Q_{\rm conv}-\frac{h-H_0}{\tau_{\rm cool}}.\label{eq:linear_sw_continuity}
\end{align}

The stochastic convective forcing is replaced below by its mean-field response to a small vorticity perturbation. The exclusion and cooldown masks are neglected, so the resulting forcing amplitude should be interpreted as a nominal estimate.

\subsection{Mean Convective Forcing}

We first average the convective forcing over a time interval $\Delta T$ and over the physical polar domain $\Omega_{\rm D}$, whose area is $A_{\rm D}=\pi R_{\rm D}^2$. If $\mathcal{N}_T$ events occur during this interval, then
\begin{align}
\overline{Q}_{\rm conv}&\equiv\frac{1}{A_{\rm D}\Delta T} \int_{\Delta T}\int_{\Omega_{\rm D}} Q_{\rm conv}\,\mathrm{d}A\,\mathrm{d}t \nonumber\\
&=\frac{\Delta h}{A_{\rm D}\Delta T}\sum_{i=1}^{\mathcal{N}_T}\int_{\Omega_{\rm D}}\exp\left[-\frac{|\mathbf{x}-\mathbf{x}_i|^2}{2R_{\rm conv}^2}\right]\mathrm{d}A\int_{t_{{\rm c},i}}^{t_{{\rm p},i}}\frac{\mathrm{d}}{\mathrm{d}t}\exp\left[-\frac{(t-t_{{\rm p},i})^2}{\tau_{\rm conv}^2}\right]\mathrm{d}t .\label{eq:mean_forcing_integral}
\end{align}

Defining
\begin{equation}
\Gamma_\tau\equiv\frac{t_{\rm p}-t_{\rm c}}{\tau_{\rm conv}},
\end{equation}
the temporal integral is $1-\exp(-\Gamma_\tau^2)$. For an averaging interval containing many events,
$\mathcal{N}_T/\Delta T\simeq N$. Moreover, because $2\pi R_{\rm conv}^2\ll A_{\rm D}$, the spatial integral can be approximated by
\begin{equation}
\int_{\Omega_{\rm D}}\exp\left[-\frac{|\mathbf{x}-\mathbf{x}_i|^2}{2R_{\rm conv}^2}\right]\mathrm{d}A
\simeq2\pi R_{\rm conv}^2,\end{equation}
consequently,
\begin{equation}
\overline{Q}_{\rm conv}
\simeq
\Delta h\,N
\frac{2R_{\rm conv}^2}{R_{\rm D}^2}
\left[1-\exp(-\Gamma_\tau^2)\right].
\label{eq:mean_convective_forcing}
\end{equation}

The horizontally averaged equilibrium thickness satisfies
\begin{equation}
0=\overline{Q}_{\rm conv}-\frac{\overline{h}-H_0}{\tau_{\rm cool}},\qquad\overline{h}=H_0+\tau_{\rm cool}\overline{Q}_{\rm conv}.
\end{equation}
We therefore define the dimensionless convective forcing strength
\begin{align}
\Pi\equiv1-\frac{\overline{h}}{H_0}&=-\frac{\tau_{\rm cool}\overline{Q}_{\rm conv}}{H_0}\nonumber\\
&=\frac{|\Delta h|}{H_0}\frac{2R_{\rm conv}^2}{R_{\rm D}^2}N\tau_{\rm cool}\left[1-\exp(-\Gamma_\tau^2)\right],\label{eq:Pi_definition}
\end{align}
where $\Delta h<0$. In the present model, $\Gamma_\tau=2$, so the final factor is $1-e^{-4}$. Under the hydrostatic and ideal-gas scaling $h\propto T$, $\Pi$ can also be interpreted as the fractional difference between the radiative and radiative--convective equilibrium thicknesses. Because $\Pi\ll1$ in the simulations, we use $\overline{h}\simeq H_0$ in the following linearization.

\subsection{Growth Rate and Instability Criterion}

We linearize about the homogeneous resting state by writing
$h=H_0+h'$, $\mathbf{V}=\mathbf{V}'$, and
$\delta'=\nabla\boldsymbol{\cdot}\mathbf{V}'$. Taking the vertical curl of
Equation~(\ref{eq:linear_sw_momentum}) and linearizing
Equation~(\ref{eq:linear_sw_continuity}) gives
\begin{align}
\frac{\partial\zeta'}{\partial t}
&=-f_0\delta',
\label{eq:linear_zeta}\\
\frac{\partial h'}{\partial t}+H_0\delta'
&=-\frac{h'}{\tau_{\rm cool}}+Q_{\rm conv}'.
\label{eq:linear_h}
\end{align}
Eliminating $\delta'$ yields
\begin{equation}
\frac{\partial}{\partial t}
\left(
\zeta'-\frac{f_0}{H_0}h'
\right)
=
-\frac{f_0}{H_0}
\left(
Q_{\rm conv}'-\frac{h'}{\tau_{\rm cool}}
\right).
\label{eq:linear_pv_equation}
\end{equation}

We next derive the perturbation of the mean convective forcing. In the homogeneous basic state, $w_0=1$. For sufficiently small perturbations, the positive-part cutoff is inactive, and
\begin{equation}
w'
=
\eta\frac{\widetilde{\zeta}'}{f_0},
\qquad
\widetilde{\zeta}'=\zeta'*G_{\rm smooth}.
\end{equation}
Linearizing the normalized event-center density,
$\lambda=Nw/\int_{\Omega_{\rm D}}w\,\mathrm{d}A$, gives
\begin{equation}
\lambda'
=
\frac{N}{A_{\rm D}}
\left(
w'-\langle w'\rangle_{\rm D}
\right),
\qquad
\langle w'\rangle_{\rm D}
\equiv
\frac{1}{A_{\rm D}}
\int_{\Omega_{\rm D}}w'\,\mathrm{d}A .
\label{eq:linearized_event_density}
\end{equation}
Thus, normalization removes the spatially uniform perturbation. For a nonzero Fourier mode with wavevector $\mathbf{K}$ and $K=|\mathbf{K}|$, $\langle w'\rangle_{\rm D}=0$. The Fourier transforms of the smoothing kernel and the convective footprint are
\begin{equation}
\widehat{G}_{\rm smooth}
=
\exp\left(-\frac{K^2l_g^2}{4}\right),
\qquad
\widehat{G}_{\rm conv}
=
2\pi R_{\rm conv}^2
\exp\left(-\frac{K^2R_{\rm conv}^2}{2}\right).
\end{equation}
Multiplying the perturbed event density by the time-integrated event amplitude
$\Delta h\,[1-\exp(-\Gamma_\tau^2)]$ and by
$\widehat{G}_{\rm conv}$ then gives
\begin{align}
\widehat{Q}_{\rm conv}'
&=
\Delta h\left[1-\exp(-\Gamma_\tau^2)\right]
\frac{2\pi R_{\rm conv}^2N}{A_{\rm D}}
\eta\frac{\widehat{\zeta}'}{f_0}
\exp\left[
-\frac{K^2l_g^2}{4}
-\frac{K^2R_{\rm conv}^2}{2}
\right]
\nonumber\\
&=
-\frac{\Pi H_0}{\tau_{\rm cool}}
\eta\frac{\widehat{\zeta}'}{f_0}
\exp\left(-\frac{K^2\ell^2}{4}\right),
\qquad
\ell^2\equiv l_g^2+2R_{\rm conv}^2,
\label{eq:linear_feedback_forcing}
\end{align}
where Equations~(\ref{eq:mean_convective_forcing}) and
(\ref{eq:Pi_definition}) have been used. The two terms in $\ell^2$ represent attenuation by vorticity smoothing and by the finite convective footprint, respectively.

For perturbations proportional to
$\exp(i\mathbf{K}\boldsymbol{\cdot}\mathbf{x}+\gamma t)$, geostrophic balance gives
\begin{equation}
\widehat{\zeta}'
=
-\frac{g'}{f_0}K^2\widehat{h}'
=
-\frac{f_0}{H_0}L_D^2K^2\widehat{h}'.
\label{eq:linear_geostrophic_relation}
\end{equation}
Substituting Equations~(\ref{eq:linear_feedback_forcing}) and
(\ref{eq:linear_geostrophic_relation}) into
Equation~(\ref{eq:linear_pv_equation}) gives
\begin{equation}
\boxed{
\gamma(K)
=
\frac{
\Pi\eta L_D^2K^2
\exp(-K^2\ell^2/4)-1
}{
\tau_{\rm cool}(1+L_D^2K^2)
}.
}
\label{eq:linear_growth_rate}
\end{equation}

The long-wave limit is stable, with
$\gamma(0)=-\tau_{\rm cool}^{-1}$, while sufficiently short waves are suppressed by the Gaussian smoothing and convective footprint. Since
$K^2\exp(-K^2\ell^2/4)$ reaches its maximum at $K=2/\ell$, an unstable band exists only if
\begin{equation}
\boxed{
\Pi\eta>
\frac{e\ell^2}{4L_D^2}.
}
\label{eq:linear_instability_condition}
\end{equation}
Thus, the vorticity-dependent convective feedback must be strong enough to overcome radiative relaxation after spatial smoothing.

\subsection{Most Unstable Scale}

Let
\[A=\Pi\eta,\qquad s=K^2.\]
The exact condition $\partial\gamma/\partial s=0$ is
\begin{equation}
A\exp\left(-\frac{\ell^2s}{4}\right)\left(1-\frac{\ell^2s}{4}-\frac{L_D^2\ell^2s^2}{4}\right)+1=0.\label{eq:exact_stationary_condition}
\end{equation}
Equivalently, with
\begin{equation}
y=\frac{K^2\ell^2}{4},\qquad\rho=\frac{4L_D^2}{\ell^2},
\end{equation}
the most unstable mode is determined by
\begin{equation}
A e^{-y_m}\left(1-y_m-\rho y_m^2\right)+1=0,\qquad0<y_m<1.\label{eq:exact_ym_condition}
\end{equation}

A closed-form approximation follows by taking $K_m\ell\ll1$:
\begin{align}
&A\left[1-\frac{\ell^2s}{2}+\left(\frac{\ell^4}{16}-\frac{L_D^2\ell^2}{4}\right)s^2+\frac{L_D^2\ell^4}{16}s^3\right]+1\simeq0\nonumber\\
&\hspace{1cm}\xrightarrow[\ L_D^2\gg\ell^2\ ]{\text{neglect }O(s^3)}1+\frac{1}{A}-\frac{\ell^2s}{2}-\frac{L_D^2\ell^2s^2}{4}\simeq0.\label{eq:stationary_approximation}
\end{align}
The positive root is
\begin{equation}
K_m^2\simeq\frac{-1+\sqrt{1+4\frac{L_D^2}{\ell^2}\left(1+\frac{1}{\Pi\eta}\right)}}{L_D^2}.\label{eq:Km_squared}
\end{equation}
The final two steps are therefore
\begin{align}
K_m&\simeq\frac{1}{L_D}\left[\sqrt{1+4\frac{L_D^2}{\ell^2}\left(1+\frac{1}{\Pi\eta}\right)}-1\right]^{1/2},\label{eq:Km}\\
R_I\equiv\frac{\lambda_m}{2}=\frac{\pi}{K_m}&\simeq\pi L_D\left[\sqrt{1+4\frac{L_D^2}{\ell^2}\left(1+\frac{1}{\Pi\eta}\right)}-1\right]^{-1/2}.\label{eq:RI_approx}
\end{align}

For the numerical diagnostics, we do not use the asymptotic
approximation. Instead, Equation~(\ref{eq:exact_ym_condition}) is
solved numerically for every unstable case, after which
\begin{equation}
\boxed{K_m=\frac{2\sqrt{y_m}}{\ell},\qquad R_I=\frac{\pi}{K_m}=\frac{\pi\ell}{2\sqrt{y_m}}.}\label{eq:RI_exact}
\end{equation}

\section{Diagnostic Methods}\label{Appendix B}

\subsection{Kinetic-Energy Spectrum and Spectral Organization}\label{Appendix B1}

The staggered velocities on the ``C" grid are first interpolated to cell centers and decomposed into radial and azimuthal components about the pole. Their azimuthal means are subtracted within radial bins of width $\Delta x$, giving the nonaxisymmetric velocity $\mathbf{u}'=(u',v')$. To suppress the discontinuity at the nominal disk boundary, we apply a window function
\begin{equation}
W(r)=\begin{cases}
1, & r\leq0.85R_{\rm D},\\
\frac{1}{2}\left[1+\cos\left(\pi\frac{r-0.85R_{\rm D}}{0.15R_{\rm D}}\right)\right], & 0.85R_{\rm D}<r<R_{\rm D},\\
0, & r\geq R_{\rm D}.
\end{cases}
\end{equation}
After removing the window-weighted residual mean, the Fourier coefficients and two-dimensional kinetic-energy power are
\begin{equation}
\widehat{u'}=\mathcal F\{W(u'-\langle u'\rangle_W)\},\qquad \widehat{v'}=\mathcal F\{W(v'-\langle v'\rangle_W)\},\qquad P_E(\mathbf{k})=\frac{|\widehat{u'}|^2+|\widehat{v'}|^2}{2(N_xN_y)^2\langle W^2\rangle}.
\end{equation}
Here, $\mathcal F$ denotes the two-dimensional discrete Fourier transform, a hat denotes a Fourier coefficient, $\mathbf{k}=(k_x,k_y)$ is the horizontal wave vector, $\langle g\rangle_W=\sum Wg/\sum W$ is the window-weighted mean for any field $g$, and $\langle W^2\rangle$ is the grid mean of $W^2$. The isotropic spectrum is obtained by summing over annular Fourier shells:
\begin{equation}
E_n=\sum_{\left||\mathbf{k}|L/(2\pi)-n\right|<1/2}P_E(\mathbf{k}),\qquad k_n=\frac{2\pi n}{L},
\end{equation}
where $n$ is the dimensionless radial shell index. The largest shell within $2\leq n\leq64$ is refined by a three-point parabolic interpolation of $\ln E_n$, giving $n_{\rm E,peak}$ and
\begin{equation}
k_{\rm E,peak}=\frac{2\pi n_{\rm E,peak}}{L},\qquad R_E=\frac{\pi}{k_{\rm E,peak}}=\frac{L}{2n_{\rm E,peak}}.
\end{equation}
A peak is classified as low-confidence if it lies at the search boundary, changes by more than $40\%$ when the search interval is changed to $3\leq n\leq64$ or $2\leq n\leq96$, differs by more than $50\%$ from the peak of the smoothed log spectrum, or is less than $1.5$ times the median spectral power within the band.

The spectral organization index $I_o$ is defined by
\begin{equation}
I_o=\frac{1}{1+\sigma_{\ln n,E}^{2}}, \qquad
\sigma_{\ln n,E}^{2}=\sum_{n=2}^{64}p_n\left[\ln\left(\frac{n}{n_{\rm E,peak}}\right)\right]^2,\qquad
p_n=\frac{E_n}{\sum_{m=2}^{64}E_m}.
\end{equation}
Here, $p_n$ is the fraction of the band-integrated kinetic energy in shell $n$, and $\sigma_{\ln n,E}$ measures the logarithmic width of the spectrum about its peak. Thus, $I_o$ approaches unity when the energy is concentrated near a single characteristic scale and decreases as the spectrum broadens.

\subsection{Vortex Identification and Characterization}\label{Appendix B2}

For vortex identification, we define the normalized PV anomaly $\chi=q/(f_0/H_0)-1$. A fine field $\chi_{\rm f}=G_{1.35\Delta x}*\chi$ and a broad field $\chi_{\rm b}=G_{10\Delta x}*\chi$ are obtained by Gaussian convolution, and their difference $\chi_{\rm DoG}=\chi_{\rm f}-\chi_{\rm b}$ is used to distinguish coherent cores from broad background structures. Here, $G_s$ denotes a Gaussian kernel with standard deviation $s$, and $*$ denotes convolution.

Local maxima of $\chi_{\rm DoG}$ separated by at least $7\Delta x$ are used as watershed seeds. Their threshold is $\max(0.18,\operatorname{med}\chi_{\rm DoG}+2.8\sigma_{\rm MAD})$, where $\operatorname{med}$ denotes the median and $\sigma_{\rm MAD}=1.4826\,\operatorname{med}|\chi_{\rm DoG}-\operatorname{med}\chi_{\rm DoG}|$ is the robust standard-deviation estimate. Watershed segmentation of $-\chi_{\rm f}$ then separates neighboring positive-PV cores.

For a watershed component $\Omega_i$, the vortex center is the PV-weighted centroid
\begin{equation}
\mathbf{x}_i=\frac{\sum_{\mathbf{x}\in\Omega_i}\mathbf{x}\,w_i(\mathbf{x})}{\sum_{\mathbf{x}\in\Omega_i}w_i(\mathbf{x})},\qquad
w_i(\mathbf{x})=\left[\max\!\left(\chi_{\rm f}(\mathbf{x})-0.08,0\right)\right]^2.
\end{equation}
Candidates are retained only if they have an equivalent radius of at least $0.35\ {\rm Mm}$, eccentricity no larger than $0.94$, solidity of at least $0.48$, sufficient PV amplitude and prominence, a maximum azimuthal velocity of at least $4\ {\rm m\,s^{-1}}$, predominantly cyclonic circulation, and a center speed smaller than $0.75$ of their maximum azimuthal velocity. These dynamical and geometrical conditions reject extrema located on elongated spiral arms. Candidates separated by less than $1.15$ times the larger equivalent-core radius are treated as subpeaks of the same core and merged, retaining the candidate with the strongest combination of PV amplitude, prominence, compactness, and low center speed.

To characterize the radial structure of each identified vortex, the azimuthally averaged tangential velocity about $\mathbf{x}_i$ is fitted by
\begin{equation}
V(r)=V_m\frac{r}{R_m}\exp\left[\frac{1-(r/R_m)^b}{b}\right],
\end{equation}
where $V_m$ is the maximum fitted tangential velocity, $R_m$ is the radius at which that maximum occurs, and $b$, known as the steepness parameter, controls the profile shape. Thus, $R_m$ separates the inner region in which the tangential velocity increases outward from the outer declining part of the vortex circulation. A fitted $R_m$ is accepted only when the diagnosed maximum lies within the sampled radial interval and the velocity-profile fit has $R^2\geq0.5$. When these criteria are not met, the equivalent-area radius $\sqrt{A_i/\pi}$, where $A_i$ is the detected core area, is used to define the excluded core in the regional diagnostic described in Appendix~\ref{Appendix B4}. Only 7 of the 953 identified vortices require this fallback.

\subsection{Structural Angular Velocity}\label{Appendix B3}

The structural rotation is diagnosed over the statistically steady interval from 30 to 40 years. At each output time, we construct
\begin{equation}
A_t(\mathbf{x})=\operatorname{clip}\left[\operatorname{sgn}(f_0)\frac{q_t(\mathbf{x})-f(\mathbf{x})/H_0}{|f_0|/H_0},0,3\right]\mathcal H\!\left(R_{\rm D}-|\mathbf{x}|\right).
\end{equation}
Here, $\operatorname{clip}(y,a,b)=\min[\max(y,a),b]$ restricts $y$ to $[a,b]$, $\operatorname{sgn}(y)$ is $1$, $0$, or $-1$ for positive, zero, or negative $y$, respectively, and $\mathcal H(s)$ is the Heaviside function, equal to unity for $s\geq0$ and zero otherwise. Thus, $A_t$ retains only the positive PV anomaly inside the nominal disk and limits exceptionally strong grid-scale peaks.

The squared-amplitude centroid is
\begin{equation}
\mathbf{x}_c(t)=\frac{\int_{\mathcal D}\mathbf{x}A_t^2(\mathbf{x})\,\mathrm dA}{\int_{\mathcal D}A_t^2(\mathbf{x})\,\mathrm dA},\qquad \mathcal D=\{\mathbf{x}:|\mathbf{x}|\leq R_{\rm D}\}.
\end{equation}
Each frame is translated to place $\mathbf{x}_c$ at the origin and interpolated onto a polar grid. The normalized correlation field is
\begin{equation}
P_t(r,\theta)\propto\sqrt{\frac{r}{R_{\rm D}}}\left[A_t(\mathbf{x}_c+r\mathbf e_r)-\left\langle A_t(\mathbf{x}_c+r\mathbf e_r)\right\rangle_\theta\right],\qquad 0.08R_{\rm D}\leq r\leq R_{\rm D},
\end{equation}
where $\mathbf e_r=(\cos\theta,\sin\theta)$, $\langle\cdot\rangle_\theta$ denotes an azimuthal mean, and the proportionality factor gives $\|P_t\|_2=1$. The factor $\sqrt{r/R_{\rm D}}$ accounts for the radial area weighting. Recentering removes translation and centroid precession, while subtracting the azimuthal mean isolates the nonaxisymmetric pattern.

For frames $i$ and $j$, the azimuthal cross-correlation is evaluated by FFT:
\begin{equation}
C_{ij}(\delta\theta)=\frac{\sum_r\mathcal F_\theta^{-1}\!\left[\widehat P_j(r,m)\widehat P_i^{\,*}(r,m)\right](\delta\theta)}{\|P_i\|_2\|P_j\|_2}.
\end{equation}
Here, $\mathcal F_\theta$ is the Fourier transform in azimuth, $m$ is the azimuthal wave number, the superscript $*$ denotes complex conjugation, and $\delta\theta$ is the trial angular displacement. The structural angular velocity is
\begin{equation}
\omega_{\rm str}=\underset{\omega}{\operatorname{arg\,max}}\ \frac{\sum_{\ell=1}^{2}\ell^{-1/2}\sum_i C_{i,i+\ell}\!\left[\omega(t_{i+\ell}-t_i)\right]}{\sum_{\ell=1}^{2}\ell^{-1/2}(N_t-\ell)},
\end{equation}
where $N_t$ is the number of analyzed frames and $\ell=1,2$ denotes adjacent and second-neighbor frame pairs. The objective is sampled at $0.25\ {\rm deg\,yr^{-1}}$ intervals and refined by parabolic interpolation, with the search range restricted so that no frame pair rotates by more than $90^\circ$. Positive $\omega_{\rm str}$ denotes counterclockwise rotation.

A nonstationary measurement is retained when the maximum correlation exceeds $0.05$, exceeds the best opposite-sign solution by at least $0.01$, has the same sign in the early and late halves of the interval, and does not occur at the search boundary. Cases with $|\omega_{\rm str}|<2\ {\rm deg\,yr^{-1}}$ and correlation above $0.05$ are classified as stationary.

\subsection{Object-Based Biot--Savart Diagnostic}\label{Appendix B4}

Using the vortex centers identified in Appendix~\ref{Appendix B2}, we construct a positive-vorticity core and, when supported by the velocity structure, an anticyclonic shield for every vortex. Let $\zeta_s=G_{\Delta x}*\zeta$ denote relative vorticity smoothed with a one-grid-cell Gaussian kernel. Watershed basins seeded near the accepted centers separate adjacent vortices out to their nearest-neighbor distances, $d_{{\rm nn},i}=\min_{j\ne i}|\mathbf x_i-\mathbf x_j|$. The core geometry is the connected component of $\zeta_s>0$ containing the seed; the integration mask $\mathcal C_i$ retains the grid cells within this component for which the unsmoothed $\zeta>0$.

The preliminary shield is the connected negative-vorticity component that touches the core and belongs to the corresponding negative-vorticity watershed basin. This topological restriction prevents the shield mask from crossing the separatrix between neighboring vortices, but it can still include part of the weak interstitial background. We therefore clip it at the first credible outer zero crossing of the azimuthal-mean tangential wind,
\begin{equation}
\overline V_{\theta,i}(r)=\frac{1}{2\pi}\int_0^{2\pi}\left[-u(\mathbf x_i+r\mathbf e_r)\sin\theta+v(\mathbf x_i+r\mathbf e_r)\cos\theta\right],\mathrm d\theta,
\end{equation}
where $\mathbf e_r=(\cos\theta,\sin\theta)$. The velocity is first smoothed over one grid cell and sampled on 360 azimuths at radial intervals of $\Delta x/2$, with the search extending to the nearest-neighbor distance. Shield candidates are allowed on both sides of the nominal-disk boundary. Where a sampling circle crosses the rectangular computational boundary, velocities are interpolated using the model's periodic continuation, so the azimuthal mean is still evaluated over a complete circle. The resulting radius clips the original watershed shield within the stored computational domain. Then starting outside the cyclonic maximum, a crossing is accepted only if the mean wind remains nonpositive for at least three radial samples and reaches a negative value whose magnitude is at least $2\%$ of the cyclonic peak before a later outer feature. If this radius is $R_{0,i}$, the accepted shield is
\begin{equation}
\mathcal S_i=\mathcal S_i^{(0)}\cap\{\mathbf x:|\mathbf x-\mathbf x_i|\leq R_{0,i}\},
\end{equation}
where $\mathcal S_i^{(0)}$ is the preliminary watershed shield. A vortex with no topologically supported shield or no credible zero crossing is retained as a core-only object. Thus,
\begin{equation}
\mathcal O_i=\begin{cases}\mathcal C_i\cup\mathcal S_i,&\mathcal S_i\ \text{is accepted},\\ \mathcal C_i,&\text{otherwise}.\end{cases}
\end{equation}
The resulting objects are disjoint by construction. The watershed boundary limits ownership between vortices, while the wind-zero clipping prevents the weak flow between them from being assigned automatically to a shield.

For target vortex $i$, the source field is divided into three disjoint classes:
\begin{equation}
\mathcal R_i^{\rm self}=\mathcal O_i,\qquad
\mathcal R_i^{\rm other}=\bigcup_{j\ne i}\mathcal O_j,\qquad
\mathcal R^{\rm bg}=\mathcal G\setminus\bigcup_j\mathcal O_j,
\end{equation}
where $\mathcal G$ is the full computational domain. The residual background is additionally separated into $\mathcal R^{\rm bg}\cap\mathcal D$ and $\mathcal R^{\rm bg}\cap\mathcal D^c$, where $\mathcal D=\{\mathbf x:|\mathbf x|\leq R_{\rm D}\}$ is the nominal disk. This last split is used only to audit the direct induction by exterior vorticity; it does not assume that every boundary influence originates outside $\mathcal D$.

For any source region $\mathcal R$, the induced velocity field is
\begin{equation}
\mathbf u^{(\mathcal R)}(\mathbf x)=\frac{1}{2\pi}\int_{\mathcal R}\zeta(\mathbf x')\frac{\hat{\mathbf z}\times(\mathbf x-\mathbf x')}{{|\mathbf x-\mathbf x'|}^2}\,\mathrm dA'.
\end{equation}
The integral is evaluated by convolution with the analytic rectangular-cell integral of the Biot--Savart kernel, treating $\zeta$ as uniform within each grid cell. The centered self-cell contribution is zero, so no arbitrary singularity cutoff is required. Rather than sampling the velocity at one center point, we use its positive-vorticity-weighted mean over the target core,
\begin{equation}
\mathbf U_i^{(\mathcal R)}=\frac{\sum_{\mathbf x_p\in\mathcal C_i}\zeta_p\mathbf u^{(\mathcal R)}(\mathbf x_p)}{\sum_{\mathbf x_p\in\mathcal C_i}\zeta_p}.
\end{equation}
With the same weights at the source and evaluation points, the double sum for the core's self-induction cancels pairwise because the kernel is antisymmetric. Hence, $\mathbf U_i^{(\mathcal R_i^{\rm self})}$ is identically the velocity induced across the core by its accepted shield; it is zero for a core-only object. Linearity provides the numerical closure
\begin{equation}
\mathbf U_i^{\rm all}=\mathbf U_i^{\rm self}+\mathbf U_i^{\rm other}+\mathbf U_i^{\rm bg},
\end{equation}
whose maximum angular-velocity closure error over the analyzed cases is $1.4\times10^{-13}\,{\rm deg\,yr^{-1}}$.

For each source class $\mathcal R$, the corresponding pattern rotation is the least-squares rigid-rotation component
\begin{equation}
\omega_{\mathcal R}=\frac{\sum_i[(\mathbf x_i-\overline{\mathbf x})\times(\mathbf U_i^{(\mathcal R)}-\overline{\mathbf U}^{(\mathcal R)})]\boldsymbol{\cdot}\hat{\mathbf z}}{\sum_i|\mathbf x_i-\overline{\mathbf x}|^2},
\end{equation}
where the overbars are arithmetic means over all identified vortices in the case. Subtracting $\overline{\mathbf U}^{(\mathcal R)}$ removes uniform translation. All identified vortices are retained in the fit, including core-only objects, and the case must contain at least two vortices. The population comparison uses the 39 cases with reliable FFT structural-drift estimates; 38 of these are nonstationary under the criterion in Appendix~\ref{Appendix B3}.

Across the 38 reliable nonstationary cases, the self-object term is clockwise, counterclockwise, or zero in 30, 7, and 1 cases, respectively, whereas the other-vortex and residual-background terms are counterclockwise and clockwise, respectively, in all 38 cases. Their median absolute angular velocities are $15.7$, $228.0$, and $196.1\ \mathrm{deg\,yr^{-1}}$. The two environmental contributions are therefore individually large but strongly opposing, while the target vortex's own asymmetric shield generally provides a smaller correction. The exterior part of the residual has a median absolute magnitude of $1.8\ \mathrm{deg\,yr^{-1}}$. This measures only exterior vorticity not assigned to the identified vortex objects.

The summed Biot--Savart estimate reproduces the observed drift direction in $35/38=92.1\%$ of the reliable nonstationary cases, correlates with the FFT structural angular velocity at $r=0.82$, and has a median magnitude ratio $|\omega_{\rm BS}/\omega_{\rm str}|=0.92$. Over these same 38 cases, the error $\omega_{\rm BS}-\omega_{\rm str}$ has a mean of $25.2\ \mathrm{deg\,yr^{-1}}$, a sample standard deviation of $61.5\ \mathrm{deg\,yr^{-1}}$, and a median absolute value of $16.5\ \mathrm{deg\,yr^{-1}}$. The diagnostic therefore has a positive mean bias and substantial case-to-case scatter. Because the two environmental terms nearly cancel, their individual amplitudes depend on the object masks. With the source field, target cores, and fitted vortex set held fixed, repartitioning changes these contributions but leaves their sum unchanged by linearity.

The calculation uses the infinite-plane relative-vorticity kernel and omits periodic-image corrections and divergent flow. Periodic continuation is used only to determine the wind-profile boundary; the induced-velocity calculation still uses the free-space kernel over the stored domain. It also compares a final-frame induced velocity with a structural angular velocity fitted over the steady interval. Moreover, a beta-gyre or damping-generated anomaly is assigned according to its realized spatial location, not its causal origin. The result should therefore be interpreted as an object-based spatial decomposition of rotational advection, rather than as a complete dynamical attribution of the drift.

\bibliography{ref}{}

@ARTICLE{Fletcher2008, title={Temperature and Composition of Saturn’s Polar Hot Spots and Hexagon}, volume={319}, url={http://dx.doi.org/10.1126/science.1149514}, DOI={10.1126/science.1149514},  number={5859}, journal={Science}, publisher={American Association for the Advancement of Science (AAAS)}, author={Fletcher, L. N. and Irwin, P. G. J. and Orton, G. S. and Teanby, N. A. and Achterberg, R. K. and Bjoraker, G. L. and Read, P. L. and Simon-Miller, A. A. and Howett, C. and de Kok, R. and Bowles, N. and Calcutt, S. B. and Hesman, B. and Flasar, F. M.}, year={2008}, month={Jan}, pages={79–81}, language={en} }

@ARTICLE{Ingersoll2020, title={Cassini Exploration of the Planet Saturn: A Comprehensive Review}, volume={216}, url={http://dx.doi.org/10.1007/s11214-020-00751-1}, DOI={10.1007/s11214-020-00751-1}, number={8}, journal={Space Science Reviews}, publisher={Springer Science and Business Media LLC}, author={Ingersoll, Andrew P.}, year={2020}, month={Oct}, language={en} }

@ARTICLE{Li2018, title={Less absorbed solar energy and more internal heat for Jupiter}, volume={9}, url={http://dx.doi.org/10.1038/s41467-018-06107-2}, DOI={10.1038/s41467-018-06107-2},  number={1}, journal={Nature Communications}, publisher={Springer Science and Business Media LLC}, author={Li, Liming and Jiang, X. and West, R. A. and Gierasch, P. J. and Perez-Hoyos, S. and Sanchez-Lavega, A. and Fletcher, L. N. and Fortney, J. J. and Knowles, B. and Porco, C. C. and Baines, K. H. and Fry, P. M. and Mallama, A. and Achterberg, R. K. and Simon, A. A. and Nixon, C. A. and Orton, G. S. and Dyudina, U. A. and Ewald, S. P. and Schmude, R. W., Jr.}, year={2018}, month={Sep}, language={en} }

@ARTICLE{ONeill2016, title={Weak Jets and Strong Cyclones: Shallow-Water Modeling of Giant Planet Polar Caps}, volume={73}, url={http://dx.doi.org/10.1175/jas-d-15-0314.1}, DOI={10.1175/jas-d-15-0314.1},  number={4}, journal={Journal of the Atmospheric Sciences}, publisher={American Meteorological Society}, author={O'Neill, Morgan E and Emanuel, Kerry A. and Flierl, Glenn R.}, year={2016}, month={Apr}, pages={1841–1855}, language={en} }

@ARTICLE{Shi2026, title={Polar vortex dynamics on gas giants: Insights from 2D energy cascades}, volume={123}, url={http://dx.doi.org/10.1073/pnas.2500791123}, DOI={10.1073/pnas.2500791123},  number={4}, journal={Proceedings of the National Academy of Sciences}, publisher={Proceedings of the National Academy of Sciences}, author={Shi, Jiaru and Kang, Wanying}, year={2026}, month={Jan}, language={en} }

@ARTICLE{Brueshaber2019, title={Dynamical regimes of giant planet polar vortices}, volume={323}, url={http://dx.doi.org/10.1016/j.icarus.2019.02.001}, DOI={10.1016/j.icarus.2019.02.001}, journal={Icarus}, publisher={Elsevier BV}, author={Brueshaber, Shawn R. and Sayanagi, Kunio M. and Dowling, Timothy E.}, year={2019}, month={May}, pages={46–61}, language={en} }

@ARTICLE{Brueshaber2021, title={Effects of forcing scale and intensity on the emergence and maintenance of polar vortices on Saturn and Ice Giants}, volume={361}, url={http://dx.doi.org/10.1016/j.icarus.2021.114386}, DOI={10.1016/j.icarus.2021.114386}, journal={Icarus}, publisher={Elsevier BV}, author={Brueshaber, Shawn R. and Sayanagi, Kunio M.}, year={2021}, month={Jun}, pages={114386}, language={en} }

@ARTICLE{
Siegelman2022-1,
author = {Lia Siegelman  and William R. Young  and Andrew P. Ingersoll },
title = {Polar vortex crystals: Emergence and structure},
journal = {Proceedings of the National Academy of Sciences},
volume = {119},
number = {17},
pages = {e2120486119},
year = {2022},
doi = {10.1073/pnas.2120486119},
URL = {https://www.pnas.org/doi/abs/10.1073/pnas.2120486119},
eprint = {https://www.pnas.org/doi/pdf/10.1073/pnas.2120486119},
}

@ARTICLE{Siegelman2022-2,
    author = {Lia Siegelman and Patrice Klein and Andrew P. Ingersoll and Shawn P, Ewald and William R. Young and Annalisa Bracco and Alessandro Mura and Alberto Adriani and Davide Grassi and Christina Plainaki and Giuseppe Sindoni} ,
    title = {Moist convection drives an upscale energy transfer at Jovian high latitudes},
    journal = {Nature Physics},
    volume = {18},
    pages = {357-361},
    year = {2022},
    doi = {https://doi.org/10.1038/s41567-021-01458-y}
}

@ARTICLE{Yadav2020, title={Deep convection–driven vortex formation on Jupiter and Saturn}, volume={6}, url={http://dx.doi.org/10.1126/sciadv.abb9298}, DOI={10.1126/sciadv.abb9298},  number={46}, journal={Science Advances}, publisher={American Association for the Advancement of Science (AAAS)}, author={Yadav, Rakesh Kumar and Heimpel, Moritz and Bloxham, Jeremy}, year={2020}, month={Nov}, language={en} }

@ARTICLE{Donnelly, title={Global climate modelling of Saturn’s atmosphere, Part V: Large-scale vortices}, volume={425}, url={http://dx.doi.org/10.1016/j.icarus.2024.116302}, DOI={10.1016/j.icarus.2024.116302}, journal={Icarus}, publisher={Elsevier BV}, author={Donnelly, Padraig T. and Spiga, Aymeric and Guerlet, Sandrine and James, Matt K. and Bardet, Deborah}, year={2025}, month={Jan}, pages={116302}, language={en} }

@ARTICLE{Dyudina2009,
title = {Saturn's south polar vortex compared to other large vortices in the Solar System},
journal = {Icarus},
volume = {202},
number = {1},
pages = {240-248},
year = {2009},
issn = {0019-1035},
doi = {https://doi.org/10.1016/j.icarus.2009.02.014},
url = {https://www.sciencedirect.com/science/ARTICLE/pii/S0019103509000670},
author = {Ulyana A. Dyudina and Andrew P. Ingersoll and Shawn P. Ewald and Ashwin R. Vasavada and Robert A. West and Kevin H. Baines and Thomas W. Momary and Anthony D. {Del Genio} and John M. Barbara and Carolyn C. Porco and Richard K. Achterberg and F. Michael Flasar and Amy A. Simon-Miller and Leigh N. Fletcher}
}

@ARTICLE{Schecter2009,
    author = {David A. Schecter and Timothy J. Dunkerton},
    title = {Hurricane formation in diabatic Ekman turbulence},
    journal = {Quarterly Journal of the Royal Meteorological Society},
    year = {2009},
    volume = {135},
    number = {641},
    page = {823-838},
    doi = {10.1002/qj.405}
}

@ARTICLE { Ooyama1969,
      author = "Katsuyuki  Ooyama",
      title = "Numerical Simulation of the Life Cycle of Tropical Cyclones",
      journal = "Journal of Atmospheric Sciences",
      year = "1969",
      publisher = "American Meteorological Society",
      address = "Boston MA, USA",
      volume = "26",
      number = "1",
      doi = "10.1175/1520-0469(1969)026<0003:NSOTLC>2.0.CO;2",
      pages=      "3 - 40",
      url = "https://journals.ametsoc.org/view/journals/atsc/26/1/1520-0469_1969_026_0003_nsotlc_2_0_co_2.xml"
}

@ARTICLE { Garner2023,
      author = "Stephen T. Garner",
      title = "TC Worlds in a Three-Level Model",
      journal = "Journal of the Atmospheric Sciences",
      year = "2023",
      publisher = "American Meteorological Society",
      address = "Boston MA, USA",
      volume = "80",
      number = "11",
      doi = "10.1175/JAS-D-22-0089.1",
      pages=      "2587 - 2611",
      url = "https://journals.ametsoc.org/view/journals/atsc/80/11/JAS-D-22-0089.1.xml"
}

@ARTICLE { FuONeill2025c,
      author = "Hao Fu and Morgan E. O’Neill",
      title = "A Quasi-2D Model of Convectively Coupled Vortices",
      journal = "Journal of the Atmospheric Sciences",
      year = "2025",
      publisher = "American Meteorological Society",
      address = "Boston MA, USA",
      volume = "82",
      number = "7",
      doi = "10.1175/JAS-D-24-0155.1",
      pages=      "1317 - 1340",
      url = "https://journals.ametsoc.org/view/journals/atsc/82/7/JAS-D-24-0155.1.xml"
}

@ARTICLE{Cai2026,
    author = {T. Cai},
    title = {Progress in research on atmospheric vortices of gas giant planets(in Chinese)},
    journal = {Reviews of Geophysics and Planetary Physics},
    year = {2026},
    doi = {10.19975/j.dqyxx.2025-041},
    volume = {57},
    number = {0},
    page = {1-13}
    
}

@ARTICLE{Li2020,
author = {Cheng Li  and Andrew P. Ingersoll  and Alexandra P. Klipfel  and Harriet Brettle },
title = {Modeling the stability of polygonal patterns of vortices at the poles of Jupiter as revealed by the <i>Juno</i> spacecraft},
journal = {Proceedings of the National Academy of Sciences},
volume = {117},
number = {39},
pages = {24082-24087},
year = {2020},
doi = {10.1073/pnas.2008440117},
URL = {https://www.pnas.org/doi/abs/10.1073/pnas.2008440117},
eprint = {https://www.pnas.org/doi/pdf/10.1073/pnas.2008440117}}

@article{Li2006,
title = {Interaction of moist convection with zonal jets on Jupiter and Saturn},
journal = {Icarus},
volume = {180},
number = {1},
pages = {113-123},
year = {2006},
issn = {0019-1035},
doi = {https://doi.org/10.1016/j.icarus.2005.08.016},
url = {https://www.sciencedirect.com/science/article/pii/S0019103505003222},
author = {Liming Li and Andrew P. Ingersoll and Xianglei Huang}
}

@article{BENZEGGOUTA2026119877,
title = {A laboratory model for Jovian polar vortex crystals},
journal = {Earth and Planetary Science Letters},
volume = {680},
pages = {119877},
year = {2026},
issn = {0012-821X},
doi = {https://doi.org/10.1016/j.epsl.2026.119877},
url = {https://www.sciencedirect.com/science/article/pii/S0012821X26000609},
author = {Djihane Benzeggouta and Benjamin Favier and Michael {Le Bars}}
}

@article{gavriel2023,
author = {Gavriel, Nimrod and Kaspi, Yohai},
title = {The Westward Drift of Jupiter's Polar Cyclones Explained by a Center-of-Mass Approach},
journal = {Geophysical Research Letters},
volume = {50},
number = {19},
pages = {e2023GL103635},
doi = {https://doi.org/10.1029/2023GL103635},
url = {https://agupubs.onlinelibrary.wiley.com/doi/abs/10.1029/2023GL103635},
eprint = {https://agupubs.onlinelibrary.wiley.com/doi/pdf/10.1029/2023GL103635},
note = {e2023GL103635 2023GL103635},
year = {2023}
}

@ARTICLE{FuONeill2025b,
  author  = {Fu, Hao and O'Neill, Morgan E.},
  title   = {Mean Cold Pool Size of Quasi-Equilibrium Convection.
             Part II: The Survival Competition Hypothesis},
  journal = {Journal of the Atmospheric Sciences},
  year    = {2025},
  volume  = {82},
  number  = {2},
  pages   = {251--265},
  doi     = {10.1175/JAS-D-24-0052.1}
}

@article{LITTLE1999306,
title = {Galileo Images of Lightning on Jupiter},
journal = {Icarus},
volume = {142},
number = {2},
pages = {306-323},
year = {1999},
issn = {0019-1035},
doi = {https://doi.org/10.1006/icar.1999.6195},
url = {https://www.sciencedirect.com/science/article/pii/S001910359996195X},
author = {Blane Little and Clifford D. Anger and Andrew P. Ingersoll and Ashwin R. Vasavada and David A. Senske and H.Herbert Breneman and William J. Borucki and  {The Galileo SSI Team}}
}

@article{GUNNARSON2023115228,
title = {Multiple convective storms within a single cyclone on Saturn},
journal = {Icarus},
volume = {389},
pages = {115228},
year = {2023},
issn = {0019-1035},
doi = {https://doi.org/10.1016/j.icarus.2022.115228},
url = {https://www.sciencedirect.com/science/article/pii/S0019103522003220},
author = {Jacob L. Gunnarson and Kunio M. Sayanagi and Georg Fischer and Trevor Barry and Anthony Wesley and Ulyana A. Dyudina and Shawn P. Ewald and Andrew P. Ingersoll}
}

@article{DYUDINA20131020,
title = {Saturn’s visible lightning, its radio emissions, and the structure of the 2009–2011 lightning storms},
journal = {Icarus},
volume = {226},
number = {1},
pages = {1020-1037},
year = {2013},
issn = {0019-1035},
doi = {https://doi.org/10.1016/j.icarus.2013.07.013},
url = {https://www.sciencedirect.com/science/article/pii/S001910351300314X},
author = {Ulyana A. Dyudina and Andrew P. Ingersoll and Shawn P. Ewald and Carolyn C. Porco and Georg Fischer and Yoav Yair}
}

@article{Fletcher2026,
author = {Fletcher, L. N. and Zhang, Z. and Brown, S. and Oyafuso, F. A. and Rogers, J. H. and Wong, M. H. and Mura, A. and Eichstädt, G. and Orton, G. S. and Brueshaber, S. and Sankar, R. and Li, C. and Levin, S. M. and Biagiotti, F. and Guillot, T. and Ingersoll, A. P. and Grassi, D. and Hansen, C. J. and Bolton, S. and Waite, J. H.},
title = {Structure of Jupiter's High-Latitude Storms: Folded Filamentary Regions Revealed by Juno},
journal = {Journal of Geophysical Research: Planets},
volume = {131},
number = {1},
pages = {e2025JE009315},
doi = {https://doi.org/10.1029/2025JE009315},
url = {https://agupubs.onlinelibrary.wiley.com/doi/abs/10.1029/2025JE009315},
eprint = {https://agupubs.onlinelibrary.wiley.com/doi/pdf/10.1029/2025JE009315},
note = {e2025JE009315 2025JE009315},
year = {2026}
}

@article{Molinari2004,
  author  = {Molinari, John and Vollaro, David and Corbosiero, Kristen L.},
  title   = {Tropical Cyclone Formation in a Sheared Environment: A Case Study},
  journal = {Journal of the Atmospheric Sciences},
  year    = {2004},
  volume  = {61},
  number  = {21},
  pages   = {2493--2509},
  doi     = {10.1175/JAS3291.1}
}

@article { Mapes1993,
      author = "Brian E.  Mapes",
      title = "Gregarious Tropical Convection",
      journal = "Journal of Atmospheric Sciences",
      year = "1993",
      publisher = "American Meteorological Society",
      address = "Boston MA, USA",
      volume = "50",
      number = "13",
      doi = "10.1175/1520-0469(1993)050<2026:GTC>2.0.CO;2",
      pages=      "2026 - 2037",
      url = "https://journals.ametsoc.org/view/journals/atsc/50/13/1520-0469_1993_050_2026_gtc_2_0_co_2.xml"
}

@article { FuYang2026,
      author = "Hao Fu and Da Yang",
      title = "Cloud Sync in Response to Wave-Like Large-Scale Forcings",
      journal = "Journal of the Atmospheric Sciences",
      year = "2026",
      publisher = "American Meteorological Society",
      address = "Boston MA, USA",
      volume = "83",
      number = "4",
      doi = "10.1175/JAS-D-24-0256.1",
      pages=      "429 - 452",
      url = "https://journals.ametsoc.org/view/journals/atsc/83/4/JAS-D-24-0256.1.xml"
}

@book{versteeg2007introduction,
  author    = {Versteeg, H. K. and Malalasekera, W.},
  title     = {An Introduction to Computational Fluid Dynamics: The Finite Volume Method},
  edition   = {2},
  publisher = {Pearson Education},
  address   = {Harlow, England},
  year      = {2007},
  isbn      = {978-0-13-127498-3}
}

@book{lomax2001fundamentals,
  author    = {Lomax, Harvard and Pulliam, Thomas H. and Zingg, David W.},
  title     = {Fundamentals of Computational Fluid Dynamics},
  series    = {Scientific Computation},
  publisher = {Springer},
  address   = {Berlin, Heidelberg},
  year      = {2001},
  doi       = {10.1007/978-3-662-04654-8},
  isbn      = {978-3-540-41607-4}
}

@article { LiWang1994,
      author = "Xiaofan  Li and Bin  Wang",
      title = "Barotropic Dynamics of the Beta Gyres and Beta Drift",
      journal = "Journal of Atmospheric Sciences",
      year = "1994",
      publisher = "American Meteorological Society",
      address = "Boston MA, USA",
      volume = "51",
      number = "5",
      doi = "10.1175/1520-0469(1994)051<0746:BDOTBG>2.0.CO;2",
      pages=      "746 - 756",
      url = "https://journals.ametsoc.org/view/journals/atsc/51/5/1520-0469_1994_051_0746_bdotbg_2_0_co_2.xml"
}

@article{Stellmach2014,
  title = {Approaching the Asymptotic Regime of Rapidly Rotating Convection: Boundary Layers versus Interior Dynamics},
  author = {Stellmach, S. and Lischper, M. and Julien, K. and Vasil, G. and Cheng, J. S. and Ribeiro, A. and King, E. M. and Aurnou, J. M.},
  journal = {Phys. Rev. Lett.},
  volume = {113},
  issue = {25},
  pages = {254501},
  numpages = {5},
  year = {2014},
  month = {Dec},
  publisher = {American Physical Society},
  doi = {10.1103/PhysRevLett.113.254501},
  url = {https://link.aps.org/doi/10.1103/PhysRevLett.113.254501}
}

@article{ingersoll2022vorticity,
  title = {Vorticity and divergence at scales down to 200 km within and around the polar cyclones of Jupiter},
  author = {Ingersoll, Andrew P. and Ewald, Shawn P. and Tosi, Federico and Adriani, Alberto and Mura, Alessandro and Grassi, Davide and Plainaki, Christina and Sindoni, Giuseppe and Li, Cheng and Siegelman, Lia and Young, William R. and Klein, Patrice},
  journal = {Nature Astronomy},
  volume = {6},
  pages = {1280--1286},
  year = {2022},
  doi = {10.1038/s41550-022-01774-0},
  url = {https://doi.org/10.1038/s41550-022-01774-0}
}

@article { Yang2013,
      author = "Da Yang and Andrew P. Ingersoll",
      title = "Triggered Convection, Gravity Waves, and the MJO: A Shallow-Water Model",
      journal = "Journal of the Atmospheric Sciences",
      year = "2013",
      publisher = "American Meteorological Society",
      address = "Boston MA, USA",
      volume = "70",
      number = "8",
      doi = "10.1175/JAS-D-12-0255.1",
      pages=      "2476 - 2486",
      url = "https://journals.ametsoc.org/view/journals/atsc/70/8/jas-d-12-0255.1.xml"
}

@article{LiIngersoll2015,
  author  = {Li, Cheng and Ingersoll, Andrew P.},
  title   = {Moist convection in hydrogen atmospheres and the frequency of {Saturn}'s giant storms},
  journal = {Nature Geoscience},
  year    = {2015},
  volume  = {8},
  number  = {5},
  pages   = {398--403},
  doi     = {10.1038/ngeo2405},
  url     = {https://doi.org/10.1038/ngeo2405}
}

@article{CHEN2025116438,
title = {Vortex crystals at Jupiter’s poles: Emergence controlled by initial small-scale turbulence},
journal = {Icarus},
volume = {429},
pages = {116438},
year = {2025},
issn = {0019-1035},
doi = {https://doi.org/10.1016/j.icarus.2024.116438},
url = {https://www.sciencedirect.com/science/article/pii/S0019103524004986},
author = {Sihe Chen and Andrew P. Ingersoll and Cheng Li}
}

@article{Brown2018,
  author  = {Brown, Shannon and Janssen, Michael and Adumitroaie, Virgil and Atreya, Sushil and Bolton, Scott and Gulkis, Samuel and Ingersoll, Andrew and Levin, Steven and Li, Cheng and Li, Liming and Lunine, Jonathan and Misra, Sidharth and Orton, Glenn and Steffes, Paul and Tabataba-Vakili, Fachreddin and Kolma{\v{s}}ov{\'a}, Ivana and Imai, Masafumi and Santol{\'i}k, Ond{\v{r}}ej and Kurth, William and Hospodarsky, George and Gurnett, Donald and Connerney, John},
  title   = {Prevalent lightning sferics at 600 megahertz near {Jupiter}'s poles},
  journal = {Nature},
  year    = {2018},
  volume  = {558},
  number  = {7708},
  pages   = {87--90},
  doi     = {10.1038/s41586-018-0156-5}
}

@article{FuDavis2025,
  author  = {Fu, Hao and Davis, Christopher A.},
  title   = {The Role of Momentum Transfer in Tropical Cyclogenesis: Insights from a Single-Column Model},
  journal = {Journal of the Atmospheric Sciences},
  year    = {2025},
  volume  = {82},
  number  = {3},
  pages   = {657--674},
  doi     = {10.1175/JAS-D-24-0069.1}
}

@article{FuSun2024,
  author  = {Fu, Hao and Sun, Shiwei},
  title   = {Vorticity skewness of finite-amplitude rapidly rotating {Rayleigh--B{\'e}nard} convection},
  journal = {Journal of Fluid Mechanics},
  year    = {2024},
  volume  = {997},
  pages   = {A3},
  doi     = {10.1017/jfm.2024.571}
}

@article{TabatabaVakili2020,
  author  = {Tabataba-Vakili, F. and Rogers, J. H. and
             Eichst{\"a}dt, G. and Orton, G. S. and
             Hansen, C. J. and Momary, T. W. and
             Sinclair, J. A. and Giles, R. S. and
             Caplinger, M. A. and Ravine, M. A. and
             Bolton, S. J.},
  title   = {Long-term tracking of circumpolar cyclones on
             {Jupiter} from polar observations with {JunoCam}},
  journal = {Icarus},
  year    = {2020},
  volume  = {335},
  pages   = {113405},
  doi     = {10.1016/j.icarus.2019.113405}
}

@article{Mura2022,
  author  = {Mura, A. and Scarica, P. and Grassi, D. and
             Adriani, A. and Bracco, A. and Piccioni, G. and
             Sindoni, G. and Moriconi, M. L. and Plainaki, C. and
             Ingersoll, A. and Altieri, F. and Cicchetti, A. and
             Dinelli, B. M. and Filacchione, G. and Migliorini, A. and
             Noschese, R. and Sordini, R. and Stefani, S. and
             Tosi, F. and Turrini, D.},
  title   = {Five Years of Observations of the Circumpolar
             Cyclones of {Jupiter}},
  journal = {Journal of Geophysical Research: Planets},
  year    = {2022},
  volume  = {127},
  number  = {9},
  pages   = {e2022JE007241},
  doi     = {10.1029/2022JE007241}
}

@article{Fu2026ColdPools,
  author  = {Fu, Hao},
  title   = {{How Do Cold Pools Influence the Size of Tropical Cyclone Embryos?}},
  journal = {Journal of the Atmospheric Sciences},
  year    = {2026},
  volume  = {83},
  number  = {8},
  pages   = {1541--1565},
  month   = aug,
  doi     = {10.1175/JAS-D-25-0165.1},
  url     = {https://doi.org/10.1175/JAS-D-25-0165.1}
}
\bibliographystyle{aasjournalv7}

\end{document}